\documentclass[fleqn,usenatbib]{mnras}

\usepackage{newtxtext,newtxmath}
\usepackage[T1]{fontenc}

\DeclareRobustCommand{\VAN}[3]{#2}
\let\VANthebibliography\thebibliography
\def\thebibliography{\DeclareRobustCommand{\VAN}[3]{##3}\VANthebibliography}

\usepackage{graphicx}   % figures
\usepackage{amsmath}    % advanced maths
\usepackage{verbatim}   % provides the "comment" environment used in the draft
\usepackage{url}        % \url in the acknowledgements
\usepackage{longtable}  % multi-page data tables (photometry / long RV lists)
\usepackage{graphicx}
\usepackage{subcaption}

\providecommand{\ion}[2]{#1\,\textsc{#2}}
\providecommand{\arcsec}{\mbox{$^{\prime\prime}$}}
\providecommand{\arcmin}{\mbox{$^{\prime}$}}

\title[Three New Spider Pulsars]{Three New Likely Spider Millisecond Pulsar Binaries and Optical Kinematic Tracers of Intrabinary Shocks}

\author[J. Strader et al.]{
Jay Strader,$^{1}$\thanks{E-mail: straderj@msu.edu}
Rebecca Kyer,$^{1}$
Laura Chomiuk,$^{1}$
Elias Aydi,$^{2}$
Peter Craig,$^{1}$
Thomas M. Do,$^{1}$
Isabella Molina,$^{1}$
\newauthor
Teresa Panurach,$^{3}$
Subhroja Roy,$^{4}$
Kirill V. Sokolovsky,$^{2}$
Samuel J. Swihart$^{5}$
and Ryan Urquhart$^{1}$
\\
$^{1}$Center for Data Intensive and Time Domain Astronomy, Department of Physics and Astronomy, Michigan State University, East Lansing, MI 48824, USA\\
$^{2}$Department of Physics \& Astronomy, Texas Tech University, Box 41051, Lubbock, TX 79409, USA\\
$^{3}$U.S. Naval Research Laboratory, 4555 Overlook Ave SW, Washington, DC 20375, USA\\
$^{4}$Department of Physics and Astronomy, University of Notre Dame, 225 Nieuwland Science Hall, Notre Dame, IN 46556, USA\\
$^{5}$Science, Systems and Sustainment Division, Institute for Defense Analyses, Alexandria, VA 22305, USA
}

\date{Accepted XXX. Received YYY; in original form ZZZ}

\pubyear{\the\year{}}

\begin{document}
\label{firstpage}
\pagerange{\pageref{firstpage}--\pageref{lastpage}}
\maketitle

\begin{abstract}

We report the discovery of three candidate spider millisecond pulsars via multi-wavelength follow-up of unassociated Fermi Large Area Telescope $\gamma$-ray sources. All have optical spectroscopy from the SOAR 4.1-m telescope and X-ray data from Swift. Two optical/X-ray sources are exceptionally strong redback candidates, sitting close to the centres of the localization ellipses of the $\gamma$-ray sources 4FGL J1020.4--5314 and 4FGL J1408.7--3747. We show that both optical sources are spectroscopic binaries with orbital periods (0.15--0.3 d) and mass functions (0.6--$0.9 M_{\odot}$) typical for redbacks. 4FGL J1020.4--5314 shows bright, very broad (full width at half maximum 2100 km s$^{-1}$) H$\alpha$ emission in all spectra obtained over two years. Remarkably, the H$\alpha$ radial velocities are offset in phase from the absorption-line velocities of the secondary, and match the signal expected for an emission source near the primary. This demonstrates, for the first time for any spider, that the H$\alpha$ emission can be associated with an intrabinary shock that wraps around the pulsar, opening up the possibility of using optical emission lines as kinematic tracers of the intrabinary shock in some redbacks. A third candidate spider is less secure in its association with the $\gamma$-ray source 4FGL J1429.8--0739, but has unusual properties, including a 30.5-hour orbital period, a potential subgiant secondary, and a high X-ray luminosity, making it worthy of further investigation.

\end{abstract}

% Select between one and six approved MNRAS keywords (please verify against the
% official MNRAS keyword list). A suggested starting set:

\section{Introduction}

Millisecond pulsars are thought to have been recycled to short periods through accretion from a companion \citep{Bhattacharya1991}. Typical millisecond pulsars in the field are fully recycled and have low-mass white dwarf companions \citep{Tauris06}. However, the discovery via the
 Fermi Large Area Telescope (LAT) that most millisecond pulsars are efficient emitters of GeV $\gamma$-rays \citep{Abdo13} has allowed for targeted radio, optical, and X-ray searches of Fermi-LAT $\gamma$-ray sources for new pulsars. These efforts have demonstrated that ``spider" millisecond pulsars with close main sequence-like companions are common \citep{Ray12,Roberts13}, making up $\gtrsim 15$--20\% of the fully recycled millisecond pulsar population with known companion types \citep{Swihart2022} in the Galactic field. Their previous rarity was entirely a selection effect, as spiders are often eclipsed in radio observations due to ionized gas associated with the secondary, as originally predicted by \citet{1991ApJ...379L..69T}.

The main classes of spiders are ``redbacks" with $\gtrsim 0.1  M_{\odot}$ secondaries and ``black widows" with lower-mass companions \citep{Roberts13,Breton2013,Strader19}, while a few ``huntsman" systems with evolved secondaries and $\sim 5$--10 d orbital periods have also been identified \citep{Strader15,Camilo16,Strader2025}. All spiders are X-ray-luminous due to intrabinary shocks between the pulsar and the companion \citep{Roberts2014}.

These millisecond pulsars are not simply a curiosity: $\gtrsim 80\%$ of pulsars with spins $< 2$ msec are spiders \citep{Swihart2022}, including the current record holder (a globular cluster redback; \citealt{Hessels2006}). The primaries in spiders are also more massive than typical neutron stars (e.g., \citealt{Strader19}), including several systems with claimed neutron star masses $> 2.3 M_{\odot}$, such as the black widow PSR J0952-0607 \citep{Romani2026} and the redback PSR J2215+5135 \citep{Linares18,Kandel20}. While important systematics can affect these mass measurements (e.g., \citealt{Clark2023}), it is plausible or even likely that the most massive and fastest spinning neutron stars reside in spiders. These extreme spin and mass properties are likely due to their evolutionary histories: spiders are thought to descend from progenitor binaries with short initial orbital periods and lengthy accretion \citep{Chen2013,Benvenuto2014,Bellm2016,Ginzburg2021,Misra2025,Sautron2026}.

As of mid-2026\footnote{see SpiderCat; https://astro.phys.ntnu.no/SpiderCAT/} there are about 90 field spiders confirmed with detection of their radio pulsations \citep{Koljonen2025}, along with many yet unconfirmed but convincing candidates. A large fraction of these were discovered through radio, X-ray, and/or optical follow-up of previously unassociated Fermi-LAT $\gamma$-ray sources, which continues to be an effective method of finding new spiders. Optical follow-up is especially valuable for redbacks and huntsman systems, which have more luminous companions that allow optical spectroscopy to help determine the physical parameters of the binary.

As part of our ongoing programme on the SOAR 4.1-m telescope to find new millisecond pulsar binaries 
associated with Fermi $\gamma$-ray sources, here we report the discovery of three new spectroscopic binaries: two nearby redbacks with compelling evidence for being spiders (4FGL J1020.4--5314 and 4FGL J1408.7--3747) and an unusual more distant candidate spider with an uncertain classification (4FGL J1429.8--0739).

\section{Data}
\label{sec:obs}

\subsection{$\gamma$-rays}
\label{sec:gammarays}

4FGL J1020.4--5314 is first present in the 4FGL Catalog \citep{Abdollahi2020}. In the 16-year Catalog, it is fit with a LogParabola model, consistent with that expected for a pulsar. Its 0.1--100 GeV flux is ($4.2\pm0.6) \times 10^{-12}$ erg s$^{-1}$ cm$^{-2}$, which corresponds to a luminosity of $L_{\gamma} = (6.1^{+6.2}_{-3.0}) \times 10^{32}$ erg s$^{-1}$ at the Gaia parallax distance of $1.1^{+0.5}_{-0.3}$ kpc (Section \ref{sec:optcounterparts}). This uncertainty includes both the count rate and distance uncertainties.

4FGL J1408.7--3747 first appeared in the 3FGL Catalog as 3FGL J1409.2--3743, where it had a large uncertainty ellipse. In the 16-year Catalog, it is fit by a LogParabola model with a  0.1--100 GeV flux of ($2.9\pm0.4) \times 10^{-12}$ erg s$^{-1}$ cm$^{-2}$, which corresponds to a luminosity of $L_{\gamma} = (4.1^{+1.5}_{-0.9}) \times 10^{32}$ erg s$^{-1}$ at the Gaia parallax distance of $1.08^{+0.15}_{-0.12}$ kpc (Section \ref{sec:optcounterparts}).

4FGL J1429.8--0739 is first present in 4FGL. In the 16-year Catalog, it is fit by a LogParabola model, and is the faintest of these sources, with a 0.1--100 GeV flux of ($1.6\pm0.2) \times 10^{-12}$ erg s$^{-1}$ cm$^{-2}$.

\subsection{X-rays}
\label{sec:X-rays}

Swift/XRT observed 4FGL J1020.4--5314 for 4.7 ksec over several epochs between 2019 Sep 7 and 2019 Oct 7. There is a single significant X-ray source within the 68\% error ellipse of the Fermi-LAT source. This source, catalogued in the Living Swift-XRT Point Source Catalogue \citep{Evans2023} as LSXPS J102030.9--531256, has a listed 0.3--10 keV count rate of $2.3^{+1.0}_{-0.8} \times 10^{-3}$ cts s$^{-1}$, equivalent to an absorbed 0.5--10 keV flux of $\sim 8 \times 10^{-14}$ erg s$^{-1}$ cm$^{-2}$ for a $\Gamma=1.7$ power law. The integrated $N_H$ along this line of sight is $N_H = 3.4 \times 10^{21}$ cm$^{-2}$ \citep{HI4PI}, however, much of this is beyond the parallax distance of 1.1 kpc, so the actual $N_H$ is lower. An assumed $N_H$ of half of the integrated value gives an unabsorbed 0.5--10 keV flux of $1.1^{+0.5}_{-0.4} \times 10^{-13}$ erg s$^{-1}$ cm$^{-2}$, equivalent to $L_X = 1.6^{+2.0}_{-0.8} \times 10^{31}$ erg s$^{-1}$ at the Gaia parallax distance of $1.1^{+0.5}_{-0.3}$ kpc. Owing to the small number of counts, no spectral fit is possible.

4FGL J1408.7--3747 was observed by Swift/XRT for 3.3 ksec over several epochs between 2019 and 2022. As for J1020, there is a single significant X-ray source within the 68\% error ellipse of the Fermi-LAT source. This source is catalogued as LSXPS J140845.3--374710, with a 0.3--10 keV count rate of $3.7^{+1.0}_{-1.2} \times 10^{-3}$ cts s$^{-1}$, equivalent to an unabsorbed 0.5--10 keV flux of $1.6^{+0.4}_{-0.5} \times 10^{-13}$ erg s$^{-1}$ cm$^{-2}$ for a $\Gamma=1.7$ power law assuming the Galactic $N_H = 7.4 \times 10^{20}$ cm$^{-2}$. This flux is equivalent to $L_X = 2.2^{1.0}_{-0.7} \times 10^{31}$ erg s$^{-1}$ at its Gaia parallax distance of $1.08^{+0.15}_{-0.12}$ kpc. There are too few counts for a spectral fit.

Swift/XRT observed 4FGL J1429.8--0739 for 3.9 ksec between 2019 and 2020. While there are no X-ray sources within the 95\% error ellipse of the Fermi-LAT source, there are two significant sources just outside it. The nearest is LSXPS J142949.4--073302, which matches a Gaia DR3 source that has a proper motion consistent with zero as well as luminous detections in WISE bands 3 and 4, making it an almost certain background active galactic nucleus (AGN). The next closest X-ray source is LSXPS J142923.7--073456, which has a count rate of $6.1^{+1.7}_{-1.4} \times 10^{-3}$ cts s$^{-1}$, equivalent to an unabsorbed 0.5--10 keV flux of $\sim 4.7 \times 10^{-13}$ erg s$^{-1}$ cm$^{-2}$ for a $\Gamma=1.7$ power law assuming the Galactic $N_H = 5.4 \times 10^{20}$ cm$^{-2}$. There are too few counts for a spectral fit. We argue below that this latter X-ray source is a plausible counterpart to the $\gamma$-ray source and present further discussion of this association in Section \ref{sec:j1429_interp}.

\subsection{Gaia Optical Counterparts}
\label{sec:optcounterparts}

For 4FGL J1020.4--5314, the closest optical source matching the location of the X-ray source, formally separated by only 1.6\arcsec, is listed in Gaia DR3 \citep{GaiaDR3} as source 5355884278935677696, with an ICRS position of (R.A., Dec.) = (10:20:31.113,--53:12:56.31), and $G=19.9$ mag. The proper motion of this source is unusually high, at $\mu = 22.5$ mas yr$^{-1}$, suggesting a high space velocity (see  discussion in  Section \ref{sec:context}). It has a $3\sigma$ parallax that gives a zeropoint-corrected \citep{Lindegren2021} distance of $1.1^{+0.5}_{-0.3}$ kpc. The optical spectroscopy and photometry discussed below strongly suggest this optical source is the counterpart both to the X-ray source and the $\gamma$-ray source.

For 4FGL J1408.7--3747, there is an optical source that is formally 0.9\arcsec\ from the X-ray source; the next nearest source is much further (8\arcsec) away. The likely counterpart to the X-ray source has a DR3 ID of 6117523959258628224 with an ICRS position of (R.A., Dec.) = (14:08:45.454, --37:47:10.00), and $G=17.5$. 
It has a well-measured parallax that gives a zeropoint-corrected distance of $1.08^{+0.15}_{-0.12}$ kpc. As for 4FGL J1020.4--5314, the optical spectroscopy and photometry discussed below indicate this optical source is the counterpart both to the X-ray source and the $\gamma$-ray source.

For the X-ray source LSXPS J142923.7--073456 potentially associated with 4FGL J1429.8--0739, there is an optical source 2.3\arcsec\ away, well within the formal 6.2\arcsec\ Swift/XRT positional uncertainty; no other Gaia source is within 15\arcsec\ of the X-ray source. This source has Gaia DR3 ID 6331220916056226944 and $G=18.4$. It is located at an ICRS position of (R.A., Dec.) = (14:29:23.642, --07:34:55.02). Its Gaia parallax only weakly constrains the distance to a zeropoint-corrected value of $3.0^{+3.7}_{-1.1}$ kpc, essentially consistent with all distances except perhaps $\lesssim$ 1 kpc. The optical data discussed below show this source is an X-ray--bright spectroscopic binary that is plausibly a spider pulsar associated with 4FGL J1429.8--0739.

\subsection{SOAR Optical Spectroscopy}
\label{sec:SOARdata}
We obtained spectroscopy of 4FGL J1020.4--5314 with the red camera of the Goodman Spectrograph \citep{Clemens04} on the SOAR 4.1-m telescope on portions of 11 nights from 2022 April 10 to 2024 Jan 22. For all spectra we used the red camera and a 400 l mm$^{-1}$ grating with wavelength coverage from $\sim 3950$ to $7900$ \AA. Most spectra were taken with a 1.2\arcsec\ longslit, giving full-width at half-maximum (FWHM) spectral resolutions of $\sim 7.3$ \AA; a few instead used a narrower 0.95\arcsec\ slit (FWHM $\sim 5.4$ \AA). Each spectrum had an exposure time of 1500 s. 

For 4FGL J1408.7--3747, we obtained spectroscopy with SOAR/Goodman on portions of 12 nights from 2022 April 10 to 2026 Feb 5. Most data were taken with the same 400 l mm$^{-1}$ setup as for 4FGL J1020.4--5314 and exposure times of 900 or 1200 s, while a small number of spectra used a 1200 l mm$^{-1}$ or 1800 l mm$^{-1}$ grating, covering 5750--7000 \AA\ or 6215--6940 \AA, respectively.

For  4FGL J1429.8--0739, the spectroscopy from SOAR/Goodman was taken between 2025 Mar 3 and 2026 Apr 24. All spectra used the 400 l mm$^{-1}$ setup and exposure times of either 1200 or 1500 s.

All spectra were reduced and optimally extracted in the standard manner using IRAF. The initial wavelength calibration used FeAr arcs taken adjacent to each set of exposures. To help correct for residual flexure or guiding issues, we made zeropoint corrections to the wavelength scale for each spectrum by fitting the telluric features using model spectra generated with {\tt TelFit} \citep{Gullikson2014}; see \citet{Dodge2024} for more details on our methodology.

The times for all measurements in this paper are listed as Barycentric Modified Julian Dates (BMJD) on the TDB system, i.e., BJD -- 2400000.5.

\begin{figure}
	\includegraphics[width=\columnwidth]{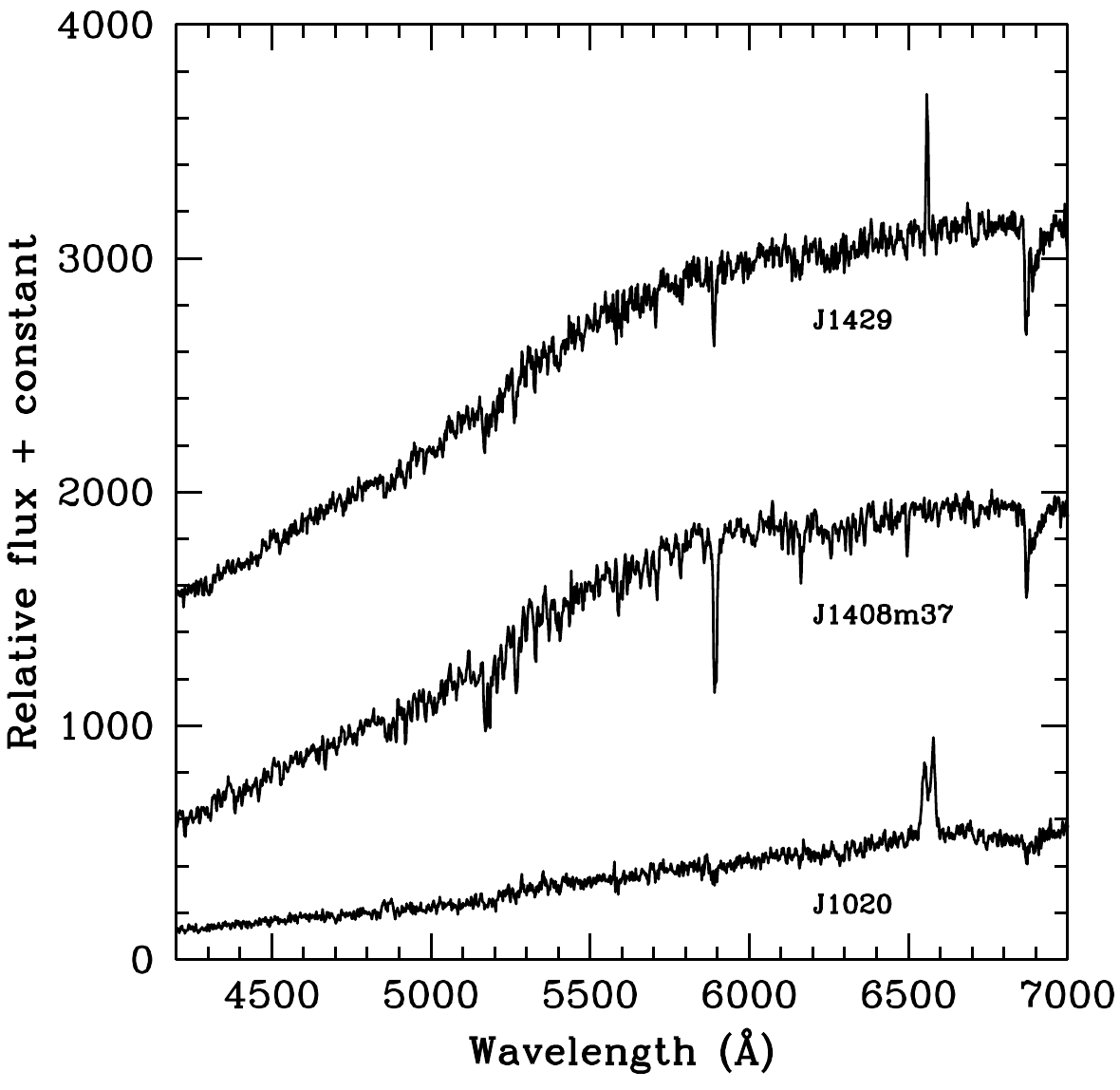}
    \caption{Sample spectra for the three binaries described in this paper. 4FGL J1020.4--5314 (bottom) shows strong broad Balmer emission with weak metal lines, 4FGL J1408.7--3747 (middle) has strong metal lines and no obvious emission, while 4FGL J1429.8--0739 (top) is intermediate between these, with weaker narrower persistent H emission.}
\label{fig:specfig}
\end{figure}

\begin{figure*}
    \centering
    \begin{subfigure}{0.49\textwidth}
        \centering
        \includegraphics[width=\linewidth]{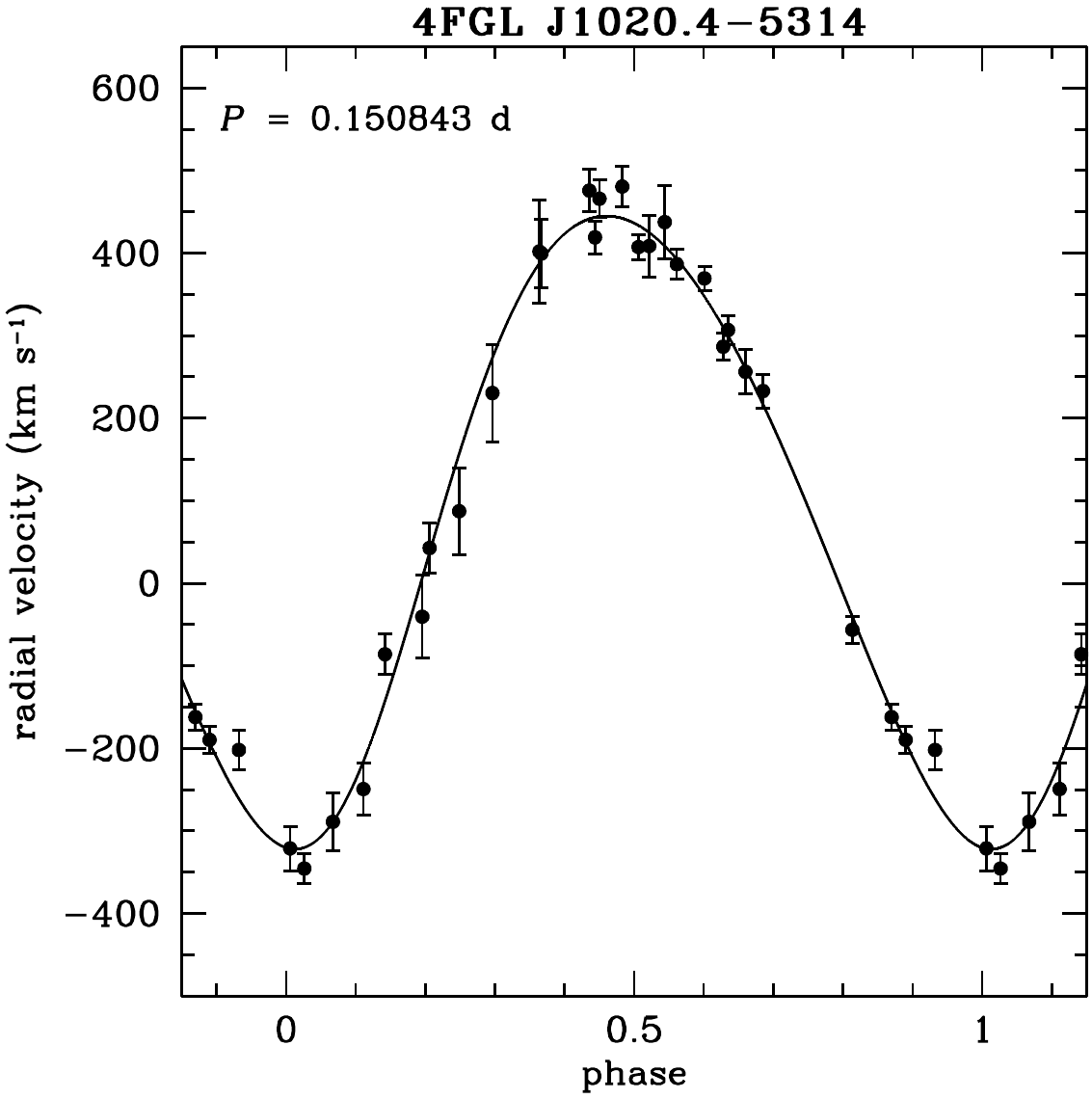}
%        \caption{}
%        \label{fig:panel_a}
    \end{subfigure}
    \hfill
    \begin{subfigure}{0.49\textwidth}
        \centering
%        \vspace{10mm}
        \includegraphics[width=\linewidth]{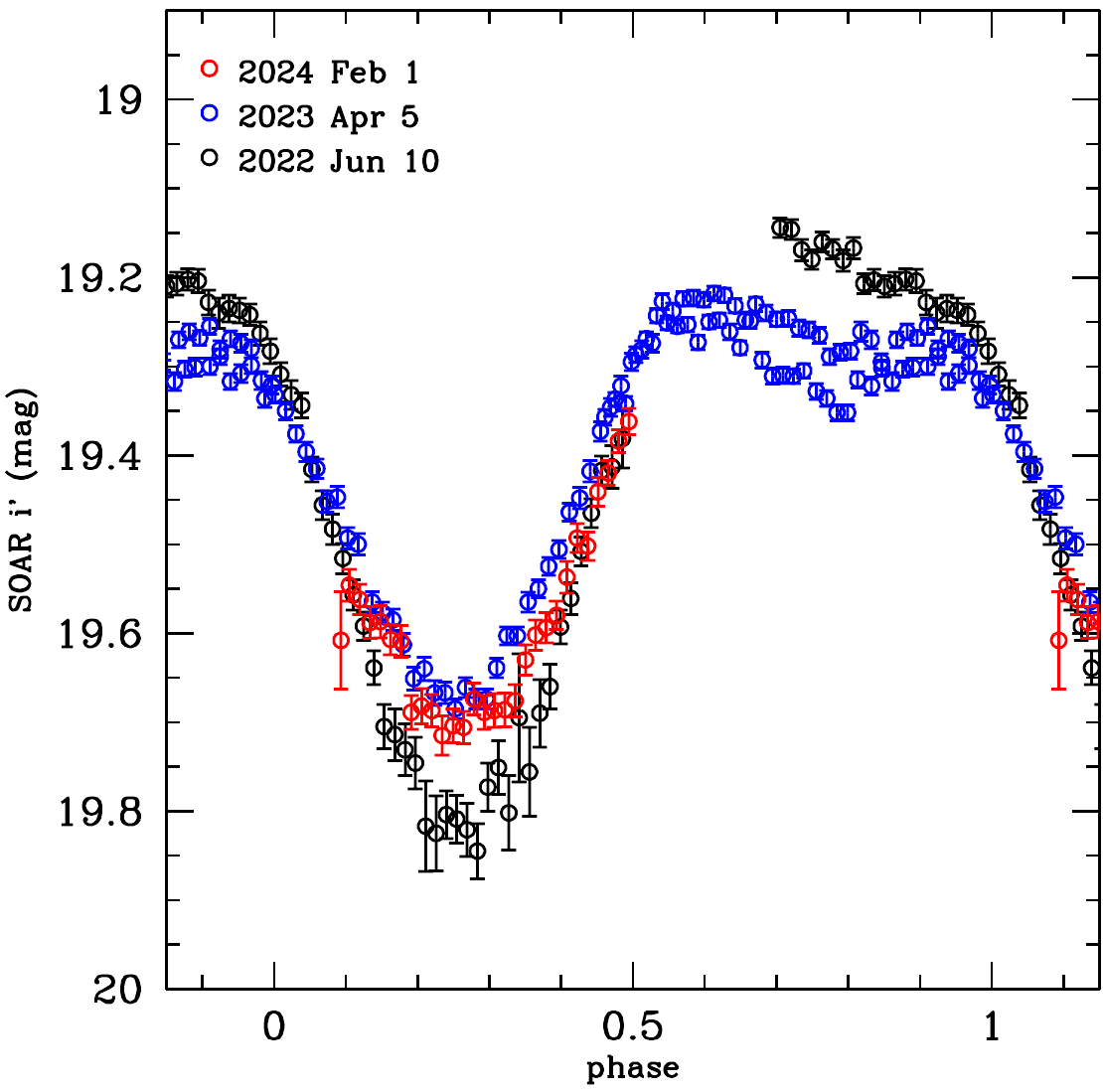}
%        \caption{}
%        \label{fig:panel_b}
    \end{subfigure}
    \caption{Panel (a). Eccentric Keplerian fit to the SOAR absorption-line radial velocities for 4FGL J1020.4--5314. Despite the improvement of this fit over a circular fit, it is almost certainly due to the effects of irradiation in a circular binary. Panel (b). Phase-folded light curve of 4FGL J1020.4--5314, covering parts of three different nights. The variability of the irradiation is evident on both short and long timescales, especially when the heated face of the secondary is visible around $\phi=0.75$.}
    \label{fig:j1020_rv}
\end{figure*}

\subsection{Optical Imaging}

We obtained time-series imaging observations of 4FGL J1020.4--5314 with SOAR/Goodman on parts of three nights (2022 Jun 9/10, 2023 Apr 5, and 2024 Feb 1), all with the red camera. These imaging observations used a SDSS $i'$ filter and an exposure time of 180 s. Seeing varied, but was typically 0.8\arcsec--1.0\arcsec.

All reduction and analysis was performed with IRAF \citep{Tody86,Tody93,Fitzpatrick24}. The reduction was done in the standard manner, using bias frames and either dome or twilight flat observations obtained the same night. For each night of observations, we did differential aperture photometry compared to a set of 50 stars within 2\arcmin\ of the target source with published $i$ photometry from SkyMapper DR4 \citep{Onken2024}. The photometry is listed in Table \ref{table:j1020_phot}. We note that the SkyMapper $i$ band is bluer than SDSS $i'$ and hence there is a colour-dependent offset between them (up to $i-i' \sim 0.1$ mag at $g-i = 2$). As we do not have colour information we have not attempted to correct for this shift, hence, our $i'$ photometry may have a small, colour-dependent systematic offset.

For 4FGL J1408.7--3747 , we obtained archival forced photometry in the $o$ filter from the ATLAS survey \citep{Tonry2018,Heinze2018} using its web service \citep{Shingles2021}, retaining all significant high-quality detections, and corrected the times to our BMJD/TDB system.

For 4FGL J1429.8--0739, there is photometry from 2018 Mar to 2025 Jul from DR24 of the Zwicky Transient Facility (ZTF; \citealt{Bellm2019}) in $g$ and $r$. As recommended in the ZTF Science Data System documentation, we removed unreliable data points (those with {\tt catflags} $> 0$). We also placed the times on the BMJD/TDB system. 

\section{Results}

Here we discuss the initial results from the optical spectroscopy and photometry of these binaries. Given the large systematic uncertainties associated with masses and inclinations derived from light curve fitting in the absence of pulsar constraints, we defer such detailed modelling until after the expected pulsars have been confirmed and timed.

\subsection{4FGL J1020.4--5314}
\label{sec:J1020_res}

The spectra (Figure \ref{fig:specfig}) show a combination of metal absorption features and strong double-peaked H$\alpha$ emission, with higher order Balmer lines and a few \ion{He}{i} lines also occasionally visible in emission.

We fit absorption-line radial velocities using {\tt RVSpecFit} \citep{Koposov2011,Koposov2019}, which performs full spectral fitting over a grid of PHOENIX \citep{Allard2016} templates convolved to the observed resolution. Regions around emission lines or telluric features were excluded from the fits, hence, in practice the fits only include metal lines. We obtained absorption line velocities for 28 spectra (Table~\ref{table:rv}), with a median uncertainty of 25 km s$^{-1}$.

Using {\tt TheJoker} \citep{Price-Whelan2017}, we initially fit a circular Keplerian model to the absorption line radial velocities. This fit was poor with large residuals, but an eccentric fit (Figure \ref{fig:j1020_rv}a) produced substantially better results. The best-fit parameters of the eccentric fit were orbital period $P = 0.1508431(2)$ d, eccentricity $e=0.111\pm0.015$, secondary semi-amplitude $K_2 = 381\pm6$ km s$^{-1}$, and systemic velocity $\gamma = 91\pm5$ km s$^{-1}$; this fit has $\chi^2$/d.o.f. = 34/22 with an r.m.s. of 28 km s$^{-1}$. The mass function implied by the radial velocity fit is $f(M) = 0.85\pm0.05 M_{\odot}$, a typical value for a spider that suggests an intermediate inclination. However, this measured value is likely biased high compared to the true value due to the irradiation of the secondary (see below).

The eccentricity of this fit is almost certainly the standard ``false" eccentricity observed in irradiated binaries where the centre of light is offset from the centre of mass \citep{Davey1992}. Heating is confirmed by variations in the effective temperature inferred from the spectral fits: the median value is around $T_{\rm eff} = 5100$ K, but with evidence for a large range from about 4850 K at inferior conjunction of the secondary to 5600 K at superior conjunction.

The light curve (Figure \ref{fig:j1020_rv}b) is heating-dominated, with a broad peak around superior conjunction of the secondary that shows evidence for being double-humped at some epochs. The amplitude of the light curve changes between the 2022 Jun 10 and 2023 Apr 5 nights from $> 0.6$ mag to 0.4--0.5 mag, respectively. In addition, on the latter night there is evidence for a smaller but still significant amplitude change within the night. Similar or even larger variations on the day side of the companion (around $\phi=0.75$) have been observed in other redbacks, including PSR J1048+2339, PSR J0838--2527, and PSR J2055+1545 \citep{Cho2018,Turchetta2025}, while the candidate redback
1FGL J0523.5--2529 shows both amplitude variations and distinctly identifiable flares \citep{Halpern2022}. Given that that the typical timescale of flares in the latter system is minutes, individual flares are perhaps not the likely cause of these amplitude variations in 4FGL J1020.4--5314. Other possible explanations discussed by \citet{Cho2018} include a change in the location or strength of the intrabinary shock, variations in the magnetic field configuration of the secondary that affects potential ducting of the pulsar wind\citep{Sanchez2017}, or changes in starspots on the secondary; all of these effects might in principle be present. The epoch of the ascending node of the pulsar inferred from the light curve is $T_0 = 60268.2755(7)$. 

%irradiation varying even within the night, perhaps due to flares as observed in other spiders (e.g., \citealt{Cho2018}). 

Despite the short orbital period of 4FGL J1020.4--5314, which would be more typical of a black widow (median 0.23 d vs.~0.39 d for redbacks; \citealt{Swihart2022}), the warm temperature of the night side of the secondary ($\sim 4850$ K) instead indicates this system is an irradiation-dominated redback; the highest night side temperatures for black widows reported in the compilation of \citet{Turchetta2023} are much cooler at $\sim 3500$ K.

Since the binary has a decently-constrained distance from its Gaia parallax, we can estimate the luminosity and radius of the secondary even in the absence of detailed light-curve modelling by using the nightside photometry and spectroscopic $T_{\rm eff}$. Using a mean nightside $i = 19.75$ mag and $T_{\rm eff} = 4850$ K, with a bolometric correction from the {\tt MIST} isochrones \citep{Dotter2016,Choi2016} and a three-dimensional $E(B-V) = 0.196$ \citep{Zucker2025}, the inferred isotropic-equivalent nightside luminosity is $L = 0.013^{+0.014}_{-0.006} L_{\odot}$ and radius $0.16^{+0.07}_{-0.04} R_{\odot}$. The parallax distance uncertainty dominates the uncertainty in these quantities. 

\subsubsection{Emission Lines}

All the spectra show double-peaked emission, most readily seen in H$\alpha$ but in a few spectra also \ion{He}{1} 5875 \AA\ just blueward of NaD (Figure \ref{fig:specfig}). The morphology varies little among the spectra. We measured the emission velocity by fitting a single Gaussian to the H$\alpha$ profile in each spectrum. Despite the double-peaked nature of the H$\alpha$ line, a single Gaussian fit is a robust route to determining the line center and width. The median FWHM of H$\alpha$ is large, about 2100 km s$^{-1}$; if a double Gaussian fit is used instead, the median peak separation is about 1200 km s$^{-1}$.

\begin{figure}[t!]
\begin{center}
	\includegraphics[width=\columnwidth]{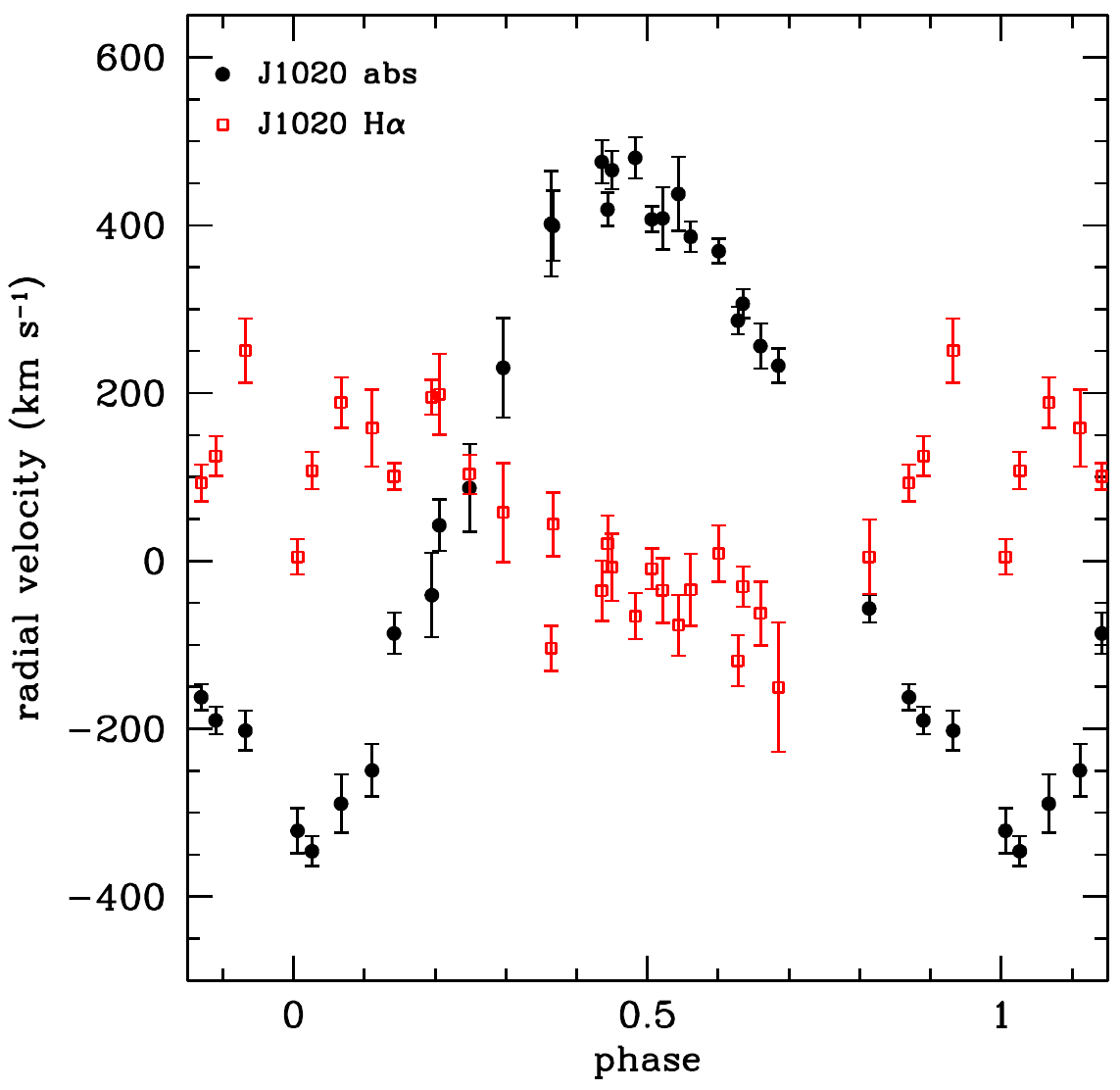}
    \caption{Comparison of absorption-line and H$\alpha$ emission line radial velocities for 4FGL J1020.4--5314. The emission line velocities are offset in phase by half an orbit and lower amplihtude, consistent with originating in an intrabinary shock wrapped around the pulsar.}
    \label{fig:j1020_em}
\end{center}
\end{figure}

The H$\alpha$ radial velocities from the single Gaussian fits are compared to the previously measured absorption-line velocities in Figure \ref{fig:j1020_em}. The H$\alpha$ velocities are \emph{anti-correlated} with the radial velocities of the secondary, offset in phase by half an orbit and of lower amplitude (a formal fit gives 88 km s$^{-1}$). This is what would be expected for an emission location near the primary rather than the secondary.

The classic interpretation of such a signature would be an accretion disc. However, there is no other evidence for an accretion disc in this system: the X-ray luminosity is only $\sim 2 \times 10^{31}$ erg s$^{-1}$ rather than the value of $\gtrsim 10^{33}$ erg s$^{-1}$ observed in transitional millisecond pulsars \citep{Linares2014,Kyer2025}, which are the only low-mass X-ray binaries that typically show $\gamma$-ray emission (see review of \citealt{Papitto2022}). In addition, there is no blue continuum in the spectra. Hence the emission must have a different source. 

In general, emission lines in spiders have been argued to originate either directly from the intrabinary shock or perhaps instead from an ionized wind from the secondary. But there has been little direct evidence to favour one scenario over the other. These H$\alpha$ kinematics, which show the emission is close to the pulsar, suggest the optical emission lines in 4FGL J1020.4--5314 originate in an intrabinary shock wrapped around the pulsar. We explore this in more detail in Section \ref{sec:eml}.

\subsection{4FGL J1408.7--3747}
\label{sec:j1408m}

\begin{figure*}
    \centering
    \begin{subfigure}{0.49\textwidth}
        \centering
        \includegraphics[width=\linewidth]{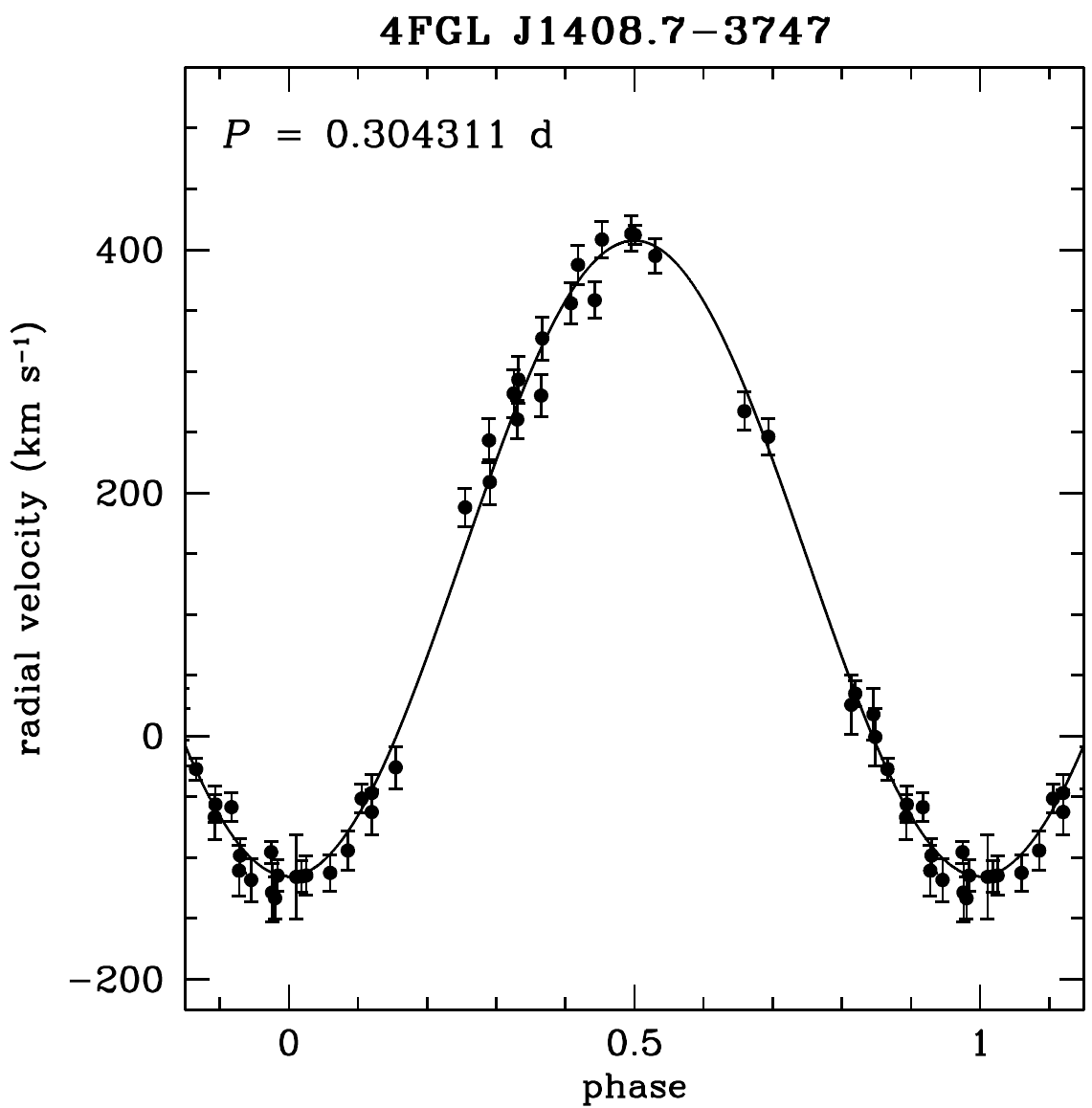}
%        \caption{}
%        \label{fig:panel_a}
    \end{subfigure}
    \hfill
    \begin{subfigure}{0.49\textwidth}
        \centering
%        \vspace{10mm}
        \includegraphics[width=\linewidth]{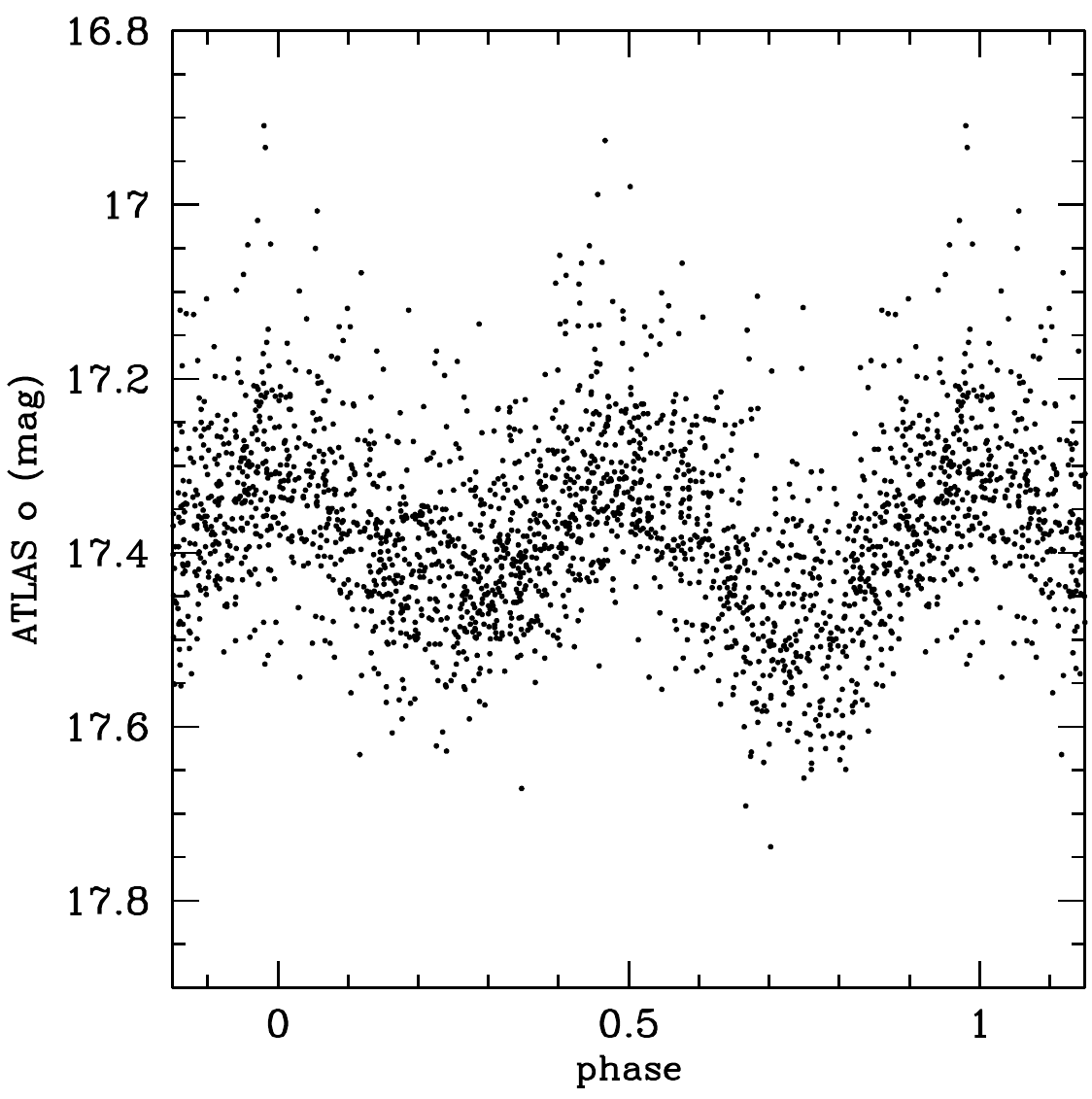}
%        \caption{}
%        \label{fig:panel_b}
    \end{subfigure}
    \caption{Panel (a). Circular Keplerian fit to the SOAR absorption-line radial velocities for 4FGL J1408.7--3747.  Panel (b). ATLAS $o$ photometry for 4FGL J1408.7--3747 phased on the best-fit spectroscopic parameters. The light curve is consistent with ellipsoidal variations with no obvious evidence of irradiation.}
    \label{fig:j1408m37_rv}
\end{figure*}

We obtained 41 usable epochs of spectroscopy from 2022 Apr 10 to 2026 Feb 5. We fit the spectra as described above for 4FGL J1020.4--5314. Unlike for that source, no emission lines are visible in the reduced spectra.

The radial velocities (Table \ref{table:j1408rv}) are well-fit by a circular orbit, with an r.m.s. scatter of 16.6 km s$^{-1}$ and $\chi^2/$d.o.f. = 43/37. This fit is shown in Figure \ref{fig:j1408m37_rv}a. The best-fit parameters are $P = 0.3043105(15)$ d, $K_2 = 262\pm4$ km s$^{-1}$, $\gamma = 146\pm3$km s$^{-1}$, and $T_0 = 60812.0799(15)$ d. The implied mass function is $f(M) = 0.56\pm0.02 M_{\odot}$. Introducing eccentricity does not meaningfully improve the fit. 

The median inferred $T_{\rm eff}$ from the spectral fits is 5725 K, and there is little evidence for variations as a function of phase the orbit, suggesting only minor irradiation. This is borne out by folding the ATLAS $o$ photometry on the spectroscopic period, which gives well-defined ellipsoidal variations (Figure \ref{fig:j1408m37_rv}b). The nightside is brighter than the dayside, as expected for a binary where gravity darkening dominates the dayside rather than irradiation. Hence this is a clear case of a redback with minimal heating.

We can also carry out a comparable calculation to that for 4FGL J1020.4--5314 to estimate the luminosity and radius of the secondary. Using $o = 17.45$ mag, $T_{\rm eff} = 5725$ K, and integrated $E(B-V) = 0.066$ \citep{Schlafly2011} gives an isotropic-equivalent nightside luminosity $0.10^{+0.02}_{-0.03} L_{\odot}$ and radius $0.32\pm0.04 R_{\odot}$. The mismatch between this M dwarf-like radius and G dwarf-like $T_{\rm eff} = 5725$ K is typical for redbacks (e.g., \citealt{Breton2013}).

\subsection{4FGL J1429.8--0739}

We obtained 63 usable spectra of this source from 2025 Mar 3 to 2026 Apr 24. These spectra were fit as described above for the two other sources, and the radial velocities are listed in Table~\ref{table:j1429rv}.

\begin{figure*}
    \centering
    \begin{subfigure}{0.49\textwidth}
        \centering
        \includegraphics[width=\linewidth]{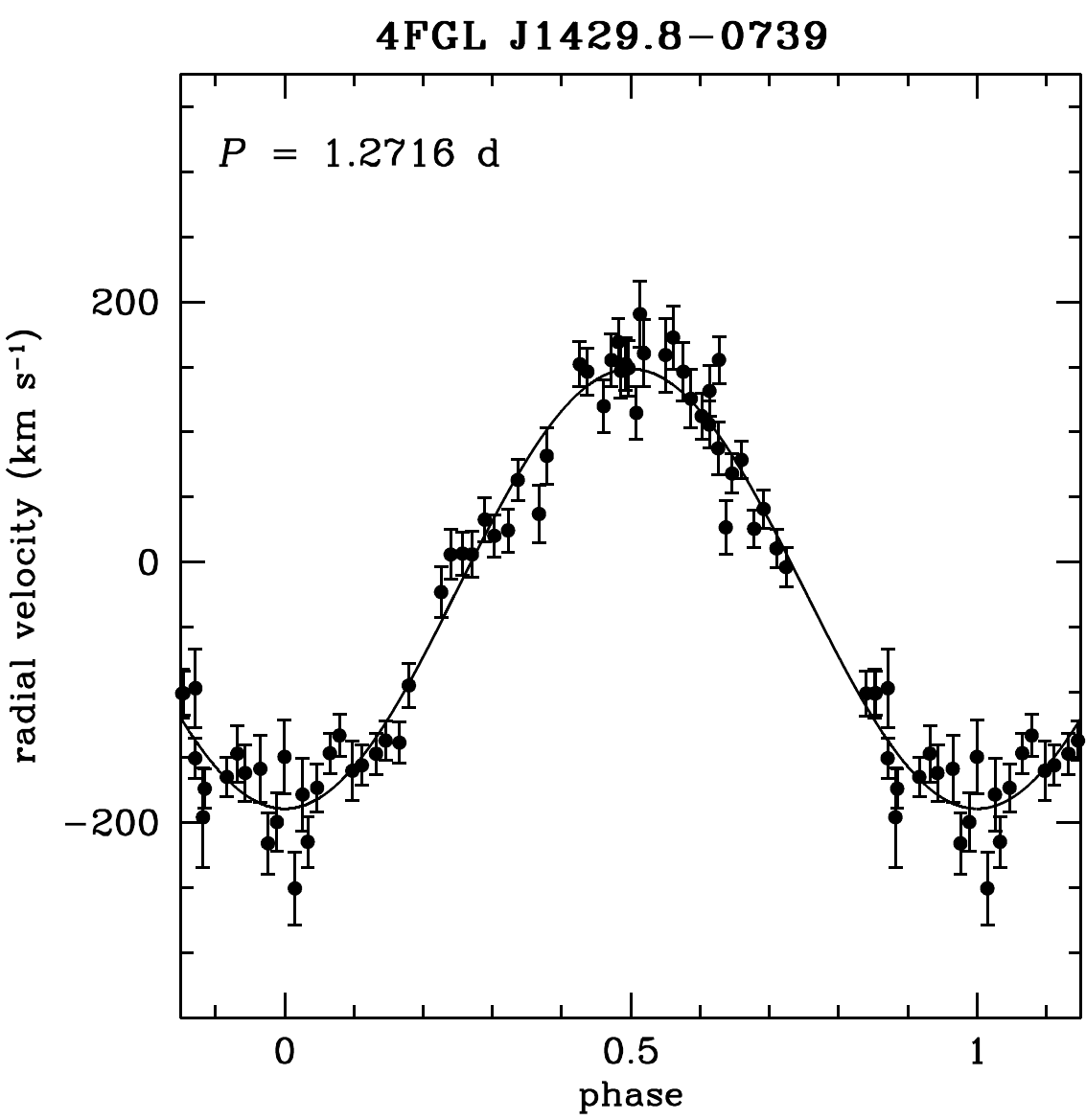}
%        \caption{}
%        \label{fig:panel_a}
    \end{subfigure}
    \hfill
    \begin{subfigure}{0.49\textwidth}
        \centering
%        \vspace{10mm}
        \includegraphics[width=\linewidth]{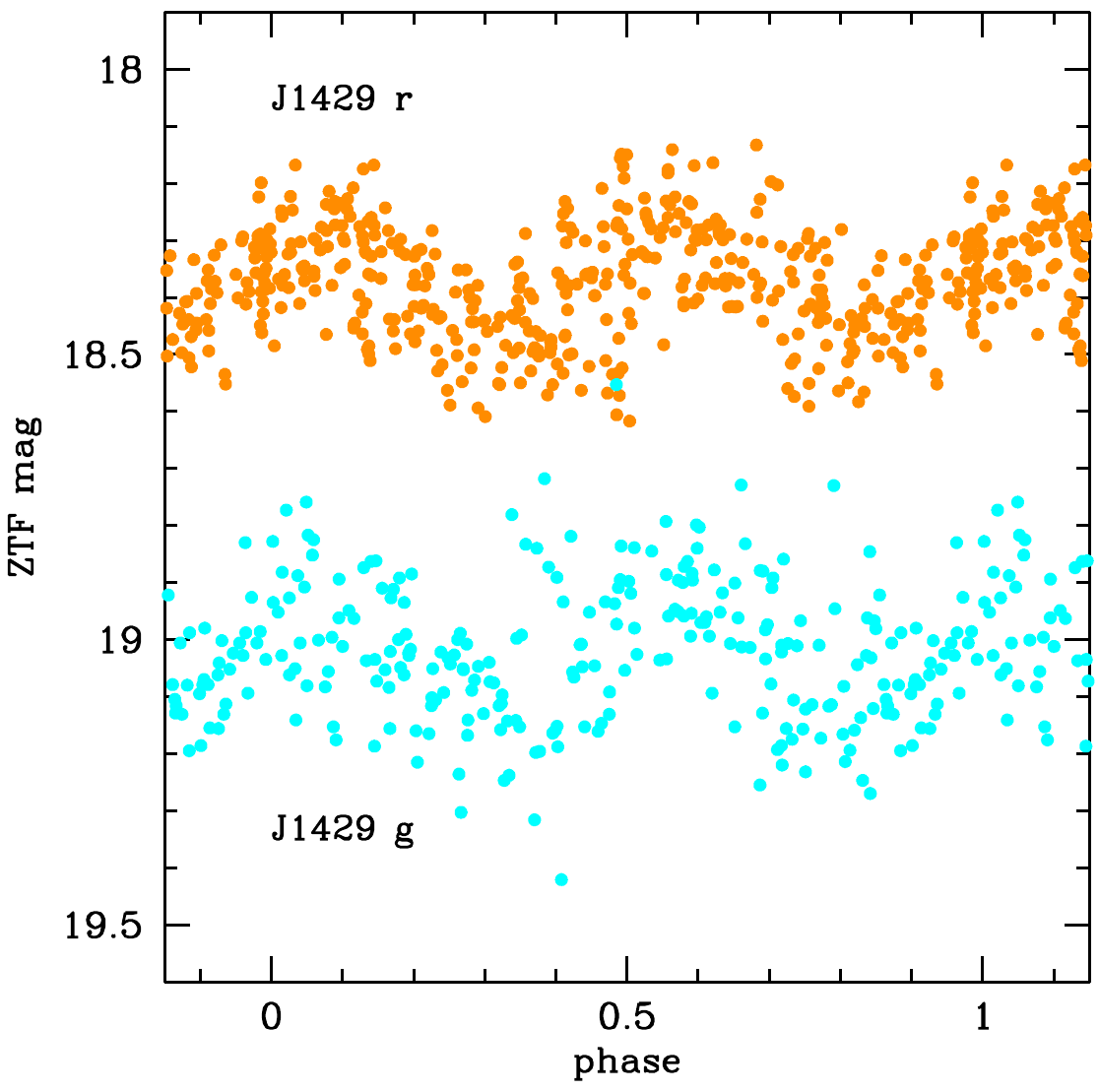}
%        \caption{}
%        \label{fig:panel_b}
    \end{subfigure}
    \caption{Panel (a). Circular Keplerian fit to the SOAR absorption-line radial velocities for 4FGL J1429.8--0739.  Panel (b). ZTF $g$ and $r$ photometry for 4FGL J1429.8--0739 phased on the best-fit spectroscopic parameters. The light curve is consistent with ellipsoidal variations with modest irradiation.}
    \label{fig:j1429a_rv}
\end{figure*}

We fit these velocities using a circular model, with best-fit parameters $P = 1.27162(7)$ d, $K_2 = 169\pm5$ km s$^{-1}$, $\gamma = -21\pm4$ km s$^{-1}$, and $T_0 = 61086.4666(56)$ d. This fit is given in Figure \ref{fig:j1429a_rv}a. The r.m.s. scatter in this fit is 26 km s$^{-1}$, with $\chi^2$/d.o.f. = 99/59. While this fit quality is not ideal, an eccentric fit is not meaningfully better. There is no posterior support for another period, and the ZTF light curve also folds well on this period. The mass function for these parameters is $f(M) = 0.64\pm0.06 M_{\odot}$.

\begin{figure}[t!]
\begin{center}
	\includegraphics[width=\columnwidth]{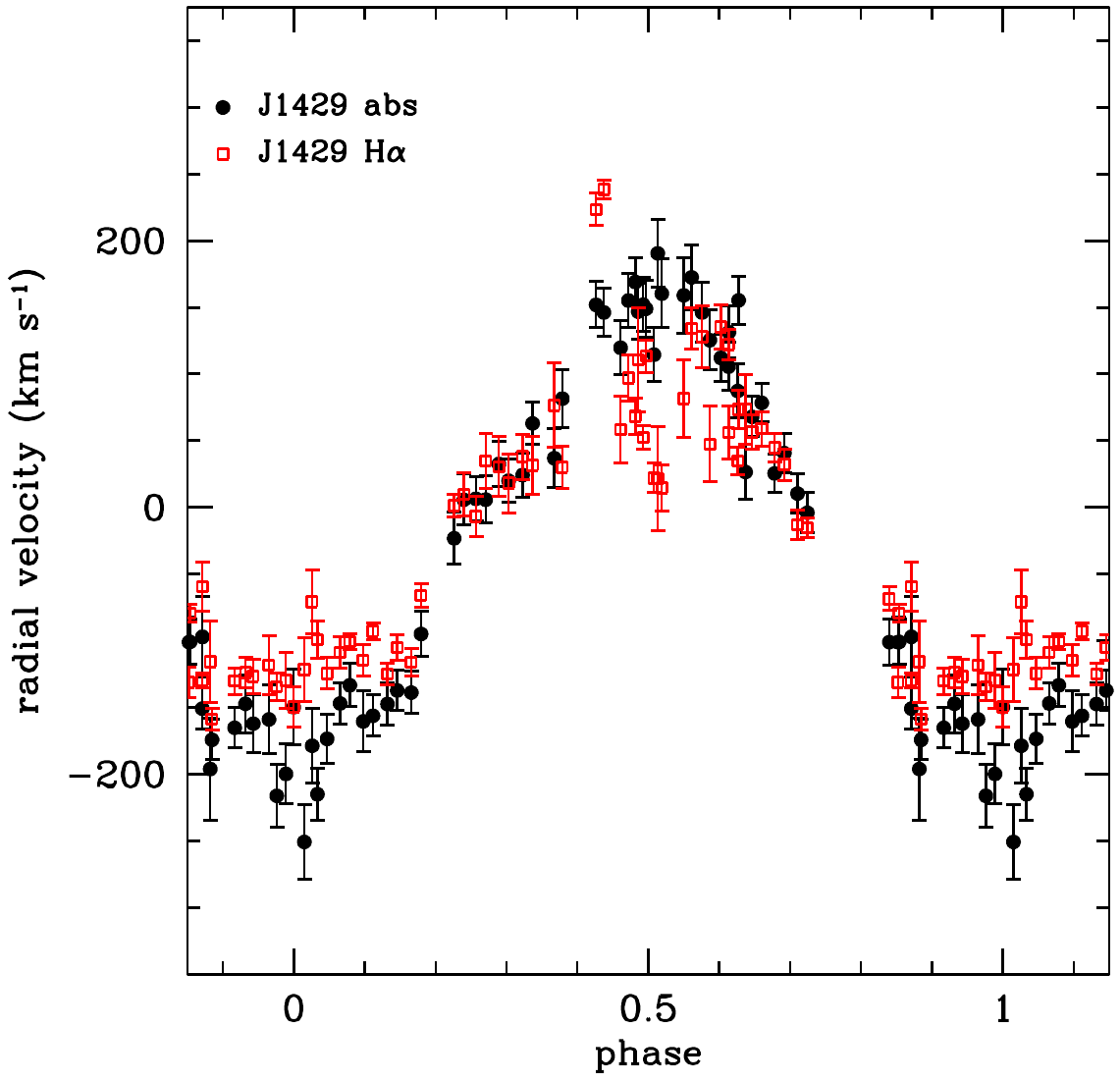}
    \caption{Comparison of absorption-line and H$\alpha$ emission line radial velocities for 4FGL J1429.8--0739. Unlike the case for 4FGL J1020.4--5314, the emission line velocities match those of the absorption lines in phase, indicating an origin in a wind or shock near the secondary.}
    \label{fig:j1429_em}
\end{center}
\end{figure}

We independently fit the radial velocities of the persistent H$\alpha$ emission line. This line is narrow, unresolved at our $\sim 300$ km s$^{-1}$ resolution in most spectra (in a few, it is clearly resolved, with FWHM up to 800 km s$^{-1}$.) We compare the absorption and emission-line velocities in Figure \ref{fig:j1429_em}. Unlike the previous result for 4FGL J1020.4--5314, these velocities largely trace the motion of the secondary in phase, though not exactly (the velocities appear flat around $\phi=0$ rather than precisely Keplerian). But the overall match shows that the H$\alpha$ emission originates near the secondary. Owing to the narrow velocity width, it seems plausible that the emission could originate in either an ionized wind or in the intrabinary shock.

In the photometry of 4FGL J1429.8--0739, we first note there is a long-term brightening of J1429 over the seven years of ZTF data that is seen in both $g$ and $r$, amounting to about $\sim 0.25$ mag. In principle, this could be due to either a $\sim 12\%$ increase in the stellar radius or half that in the equivalent $T_{\rm eff}$; since the colour does not appreciably change with time, we favour a change in radius. We note a long-term dimming was observed in the redback PSR J2129--0429 \citep{Bellm2016} and there is a possible long-term trend in the light curve of redback PSR J1723--2837 as well \citep{vanStaden2016}. Both of these binaries have unusually long periods for spiders (about 15 hr), like 4FGL J1429.8--0739. This could imply that these long-term variations are only detectable in wider binaries with more massive secondaries where the effects of heating are muted.

To allow a clear view of the orbital signal, we fit a linear model to each light curve to remove the top-level trend. Folding these detrended data on the spectroscopic orbital period and $T_0$ produces a clear ellipsoidal signal (Figure \ref{fig:j1429a_rv}b). Unlike for 4FGL J1408.7--3747, the dayside of the secondary is brighter than the nightside, indicating the presence of some heating of the dayside, as perhaps expected given the H$\alpha$ emission. The median $T_{\rm eff}$ from the spectroscopic fits is 5960 K, ranging from about 5800 K to 6100 K.

In the folded light curve the photometric minima (corresponding to the conjunctions) appear slightly offset from the expected phases of $\phi = 0.25$ and $\phi = 0.75$. Because of the offset in time between the ZTF and SOAR data (nearly 5 years in the median), the uncertainty in the spectroscopic orbital period corresponds to a substantial phase uncertainty of $\Delta \phi \sim 0.1$ at the median ZTF epoch. We independently estimated the photometric period using the Fast Periodicity Weighting algorithm \citep{Finkbeiner2025}, finding a clear signal at $P=1.27149$ or $1.27151$ d depending on the number of phase bins chosen. These are consistent with the spectroscopic period ($P = 1.27162(7)$ d) within its errors. The light curve uncertainties are large enough that we interpret this as modest but not conclusive evidence for a period in the lower end of the range consistent with the spectroscopic data.

In any case, the orbital period of 4FGL J1429.8--0739 is exceptional---longer than any redback (30.5 hr compared to 26.3 for PSR J1306--40; \citealt{Keane2018,Linares2018}). It is not even clear whether PSR J1306--40 is better classified as a redback rather than a huntsman-type spider with an evolved secondary, which also suggests uncertainty about the evolutionary state of the secondary in 4FGL J1429.8--0739. The $r$-band peak-to-trough ellipsoidal amplitude of $\sim 0.15$ -- 0.2 mag (Figure \ref{fig:j1429a_rv}b) suggests that it must be filling a substantial fraction of its Roche lobe ($\gtrsim 0.8$; \citealt{Gomel2021}). At its long orbital period, this would imply a low mean density, as typical for an evolved subgiant donor (it cannot be a red giant due to the high $T_{\rm eff}$). On the other hand, the source is faint ($r \sim 18.4$), and with little foreground reddening in this direction, it would need to be distant (likely $\gtrsim 8$ kpc) to be as bright as expected for a subgiant. The small  parallax and the faintness of the $\gamma$-ray source are at least broadly consistent with such a large distance, but it would be nevertheless unexpected. The discussion of the nature of this system is continued in Section \ref{sec:j1429_interp}.

\section{Discussion and Conclusions}
\label{sec:discussion}

\subsection{The candidate redbacks 4FGL J1020.4--5314 and 4FGL J1408.7--3747 in context}
\label{sec:context}

First, we briefly review why we believe these two sources are almost certain to be binary millisecond pulsars with low-mass H-rich secondaries. They cannot be active binaries: if the primaries were main sequence stars of the masses implied by the radial velocities of the secondaries, the primaries would dominate the optical light; instead they are not seen. The X-ray luminosities are also at least two orders of magnitude above the coronal saturation limit \citep{Vilhu1987}. Hence the primaries must be compact objects. The lack of a blue continuum in either source, or emission lines for 4FGL J1408.7--3747, excludes accreting white dwarfs. In principle a non-accreting white dwarf with an M dwarf companion could have a mass function and X-ray luminosity similar to 4FGL J1408.7--3747. However, known systems in this class (e.g., AR Sco; \citealt{Marsh2016}) have a secondary effective temperature close to that expected for a main sequence star of the secondary's stellar mass, unlike the large mismatch seen for 4FGL J1408.7--3747 (Section \ref{sec:j1408m}).

By contrast, the multi-wavelength properties of these two sources are as observed for spiders: they are the only detected X-ray sources located within the 68\% localization ellipse for curved non-variable Fermi-LAT sources and hence the most likely counterparts to these $\gamma$-ray sources. The large amplitude, time-variable optical heating seen in 4FGL J1020.4--5314 is not shown by any class of compact binary other than spiders. 4FGL J1408.7--3747 has a nightside $T_{\rm eff} = 5725$ K (typical of a G dwarf) but a size of only $0.32\pm0.04 R_{\odot}$ (typical of an M dwarf). Such discrepancies between the $T_{\rm eff}$ observed and the value predicted for a normal main sequence star of the observed radius are common for redbacks (e.g., \citealt{Breton2013,Li2014}), likely due to the stripped and/or heated nature of the donor stars (e.g., \citealt{Chen2013}). Finally, for both of the candidates, the mass functions of the primary are consistent with other redbacks \citep{Strader19}, implying neutron star-like masses for the invisible primaries. For these reasons we believe it is nearly certain that these two binaries are indeed spiders.

Both 4FGL J1020.4--5314 and 4FGL J1408.7--3747 have 0.5--10 X-ray luminosities $L_X \sim 2 \times 10^{31}$ erg s$^{-1}$, on the lower end but well within the range previously found for redbacks \citep{Urquhart2020,Koljonen2023}. Their X-ray (0.5--10 keV) to $\gamma$-ray (0.1-100 GeV) flux ratios are 0.025 and 0.055, respectively, which are values typical for redbacks \citep{Miller2020}. 

Using the Gaia proper motion and parallax distance for 4FGL J1020.4--5314, we infer a tangential velocity of $112^{+54}_{-27}$ km s$^{-1}$ and a three-dimensional velocity with respect to the local standard of rest of $131^{+45}_{-21}$ km s$^{-1}$. We also calculated the ``potential kick velocity" of the system at birth using the methodology of \citet{Atri2019}, finding a value of $142^{+24}_{-26}$ km s$^{-1}$. This value is close to the median of potential kick velocities of a sample of 14 redbacks (147 km s$^{-1}$) calculated by \citet{ODoherty2023}. For 4FGL J1408.7--3747, the tangential velocity is $48^{+7}_{-5}$ km s$^{-1}$ and the three-dimensional velocity $159^{+4}_{-3}$ km s$^{-1}$, with a potential kick velocity $179^{+22}_{-32}$ km s$^{-1}$. This value is on the higher end of, but still fully consistent with, the \citet{ODoherty2023} redback sample. Hence we find both 4FGL J1020.4--5314 and 4FGL J1408.7--3747 have kinematics typical of known redbacks.

Both of these redbacks are nearby ($\sim 1$ kpc). Another newly discovered strong candidate redback, 4FGL J2249.4+6229 \citep{Karpova2026}, is also very nearby ($\lesssim 600$ pc). This demonstrates that that the new $\gamma$-ray sources added in recent catalogues do not solely represent more distant sources: they also contain nearby sources with intrinsically lower $\gamma$-ray and/or X-ray luminosities that are nevertheless still amenable to detailed follow-up.

The basic information for the candidate spiders in this paper is shown in Table \ref{table:sources}.

\subsection{Interpreting 4FGL J1429.8--0739}
\label{sec:j1429_interp}

The case of 4FGL J1429.8--0739 is rather different than the other two redbacks considered here. The first issue is that the potential X-ray/optical binary counterpart is outside the formal 95\% positional uncertainty ellipse of the $\gamma$-ray source. This is not necessarily a fatal objection, since there is no candidate X-ray counterpart within the 95\% ellipse, and more broadly, published Fermi-LAT 95\% error ellipses have been shown to contain fewer than 95\% of true source locations when additional data became available (e.g., \citealt{Ballet2026}). 

Even though the alternative potential X-ray counterpart discussed above (a background AGN; see Section \ref{sec:X-rays}) is slightly closer to the centre, the $\gamma$-ray properties of 4FGL J1429.8--0739 in the 16-year Catalog are more similar to a pulsar than an AGN: the best-fit model is a LogParabola with significant curvature and a peak energy of $1.4\pm0.3$ GeV, typical of pulsars, though the spectral parameters are not precisely determined because the source is faint. In addition, 4FGL J1429.8--0739 shows no evidence for $\gamma$-ray variability, as expected for a pulsar but unlike typical Fermi-LAT AGN \citep{Abdollahi2020}.  Another possibility is that the single catalogued $\gamma$-ray source is actually a blend of two fainter $\gamma$-ray sources that are too close to be distinguished in existing data, but which if properly split could be separately associated with these two different plausible counterparts to the current single catalogued $\gamma$-ray source.

The second potential issue is the unusual inferred high-energy properties of the binary. At its poorly-measured X-ray flux with an assumed distance of 8 kpc, it would be the most X-ray-luminous non-accreting spider known at $L_X \sim 4\times 10^{33} (d/8 {\rm \, kpc})^2$ erg s$^{-1}$. The only source potentially in the same $L_X$ neighbourhood is the huntsman PSR J1417--4402, which has $L_X = 2.1 \times 10^{33}$ erg s$^{-1}$ (converting the X-ray flux measurement from \citealt{Swihart18} to 0.5--10 keV, and using the Gaia DR3 parallax distance of 4.2 kpc). This physical comparison is perhaps encouraging if 4FGL J1429.8--0739 does indeed have an evolved secondary like PSR J1417--4402, potentially placing it in the huntsman category. This secondary could have a strong wind that allows a larger fraction of the pulsar wind to be radiated at a shock. As mentioned above, the 1.3 d period of 4FGL J1429.8--0739 is also close to that of the spider PSR J1740--5340A in the globular cluster NGC 6397, which may have an evolved secondary \citep{Mucciarelli2013}, though this system has only $L_X \sim 10^{31}$ erg s$^{-1}$ \citep{Bogdanov2010}.

While the ideal confirmation of the nature of 4FGL J1429.8--0739 would come from the detection of a pulsar, this may be challenging, as the pulsar would be both distant and likely to be frequently eclipsed. Instead, an X-ray light curve that could show the expected orbital modulation of a spider might be the most promising route to further explore the nature of this binary.

\subsection{Emission Lines as Kinematic Tracers of Intrabinary Shocks in Spiders}
\label{sec:eml}

Optical emission lines are common in spiders, with nearly half of systems with optical spectroscopy at least sometimes showing them, typically traced by H$\alpha$ \citep{Swihart2022}. However, these emission lines are typically variable and often disappear.

Binaries with \emph{persistent} strong emission lines are rare. The redback PSR J1048+2339 \citep{Strader19,Simpson2025} appears to nearly always show broad H$\alpha$, and H$\alpha$ emission is likewise ubiquitous in the huntsman system 1FGL
J1417.7-4407 \citep{Strader15,Swihart18} and perhaps in the 1.3d orbital period spider PSR J1740--5340A in the globular cluster NGC 6397 \citep{Sabbi2003}.

In this context, finding two new candidate spiders that show emission lines in literally all spectra obtained over a long timespan is remarkable and useful. The stable profiles allow for reliable kinematics of the emission lines to be measured at all orbital phases. For 4FGL J1020.4--5314, this has allowed us to demonstrate, for the first time for any spider, that the H$\alpha$ emission is associated with a shock located near the pulsar. There is already evidence from modelling X-ray light curves of redbacks that the shock often wraps around the pulsar (e.g., \citealt{Wadiasingh2017,Kandel2019}), but of course kinematic measurements from these X-ray data are not available. Joint modelling of a future X-ray light curve for 4FGL J1020.4--5314 with optical emission line data should allow an improved understanding of the details of the intrabinary shock. 

The case of 4FGL J1429.8--0739, where the H$\alpha$-emitting gas is located close to the companion, appears more typical among spiders. Similar conclusions were reached for the black widow PSR J1311--3430 \citep{Romani2015} and the huntsman 1FGL J1417.7--4407 \citep{Swihart18}, though it was not clear whether the emitting material was associated directly with a wind from the secondary or from the shock itself.

Many candidate or confirmed spiders lack extensive spectroscopy, even for systems that have shown strong emission lines in the limited data available (e.g., 4FGL J0935.3+0901; \citealt{Wang2020}). The observational requirements to obtain emission line velocities are less stringent than for absorption lines. Hence there are likely more spiders for which observational constraints are possible on the location of the optical line-emitting gas.

\begin{table*}
\centering
\caption{Properties of the three spider millisecond-pulsar binary candidates.}
\label{table:sources}
\begin{tabular}{lcccllll}
\hline
Source & R.A. & Dec. & $G$ & $d$ & $L_{\rm X, 0.5-10}$ & $L_{\gamma, 0.1-100}$ & $P_{\rm orb}$ \\
 & (ICRS) & (ICRS) & (mag) & (kpc) & (erg\,s$^{-1}$) & (erg\,s$^{-1}$) & (d) \\
\hline
4FGL J1020.4$-$5314 & 10:20:31.113 & --53:12:56.31 & 19.9 & $1.1^{+0.5}_{-0.3}$    & $(1.6^{+2.0}_{-0.8}) \times10^{31}$          & $(6.1^{+6.2}_{-3.0}) \times10^{32}$ & 0.1508431(2)  \\
4FGL J1408.7$-$3747 & 14:08:45.454 & --37:47:10.00 & 17.5 & $1.08^{+0.15}_{-0.12}$ & $(2.2^{+1.0}_{-0.7})\times10^{31}$          & $(4.1^{+1.5}_{-0.9})\times10^{32}$ & 0.3043105(15) \\
4FGL J1429.8$-$0739 & 14:29:23.642 & --07:34:55.02 & 18.4 & \dots    & $(4\pm1)\times10^{33} (d/8$kpc)$^2$    & $10^{34} (d/8$kpc)$^2$ & 1.27162(7)    \\
\hline
\end{tabular}
\\[2pt]
\end{table*}

\subsection{Concluding Remarks}

We have presented the discovery and initial characterization of three candidate spider millisecond pulsar binaries found through optical spectroscopic follow-up of Fermi-LAT $\gamma$-ray sources. Two of these, 4FGL J1020.4--5314 ($P = 3.6$ hr) and 4FGL J1408.7--3747 ($P = 7.3$ hr), are nearby redbacks. The former is a heating-dominated system with significant and variable irradiation, while the latter shows ellipsoidal variations with no evidence for heating. The third binary, potentially associated with 4FGL J1429.8--0739, would have the longest orbital period of any known redback (30.5 hr), and may instead harbor an evolved subgiant secondary. The orbital ephemerides presented here should aid targeted radio searches for the pulsars in all three systems. 

The clear association of the H$\alpha$ emission lines in 4FGL J1020.4--5314 with an intrabinary shock located near the pulsar is in contrast to a few previous spiders, where the optical emission lines were instead located close to the companion. This suggests that modelling these emission lines in conjunction with other data (such as X-ray light curves) could be a powerful tool for understanding the dynamics of shocks in spiders.

\section*{Acknowledgements}

We acknowledge helpful comments from an anonymous referee.

We acknowledge support from NSF grant AST-2205550 and NASA grants 80NSSC23K0497, 80NSSC25K7334, 80NSSC26K0601, 80NSSC26K0226 and 80NSSC26K0658.

Based on observations obtained at the Southern Astrophysical Research (SOAR) telescope, which is a joint project of the Minist\'{e}rio da Ci\^{e}ncia, Tecnologia e Inova\c{c}\~{o}es (MCTI/LNA) do Brasil, the US National Science Foundation's NOIRLab, the University of North Carolina at Chapel Hill (UNC), and Michigan State University (MSU).

We acknowledge the use of public data from the Swift data archive.

This work has made use of data from the European Space Agency (ESA) mission {\it Gaia} (\url{https://www.cosmos.esa.int/gaia}), processed by the {\it Gaia} Data Processing and Analysis Consortium (DPAC, \url{https://www.cosmos.esa.int/web/gaia/dpac/consortium}). Funding for the DPAC has been provided by national institutions, in particular the institutions participating in the {\it Gaia} Multilateral Agreement.

%%%%%%%%%%%%%%%%%%%%%%%%%%%%%%%%%%%%%%%%%%%%%%%%%%
\section*{Data Availability}
% NOTE: A Data Availability Statement is required for all MNRAS articles.
The radial velocity and new photometric measurements underlying this article are provided in the tables in the article. The original optical images and spectra are available in the NOIRLab archive.

%%%%%%%%%%%%%%%%%%%%%%%%%%%%%%%%%%%%%%%%%%%%%%%%%%
%%%%%%%%%%%%%%%%%%%%   REFERENCES   %%%%%%%%%%%%%%%
% Uses the MNRAS BibTeX style. Keep your bibliography database (report.bib).
\bibliographystyle{mnras}
\bibliography{report}

@INCOLLECTION{Tauris06,
       author = {{Tauris}, T.~M. and {van den Heuvel}, E.~P.~J.},
        title = "{Formation and evolution of compact stellar X-ray sources}",
    booktitle = {Compact stellar X-ray sources},
         year = 2006,
       volume = {39},
        pages = {623-665},
       adsurl = {https://ui.adsabs.harvard.edu/abs/2006csxs.book..623T}
}

@ARTICLE{Abdo13,
       author = {{Abdo}, A.~A. and {Ajello}, M. and {Allafort}, A. and {Baldini}, L. and {Ballet}, J. and {Barbiellini}, G. and {Baring}, M.~G. and {Bastieri}, D. and {Belfiore}, A. and {Bellazzini}, R. and {Bhattacharyya}, B. and {Bissaldi}, E. and {Bloom}, E.~D. and {Bonamente}, E. and {Bottacini}, E. and {Brandt}, T.~J. and {Bregeon}, J. and {Brigida}, M. and {Bruel}, P. and {Buehler}, R. and {Burgay}, M. and {Burnett}, T.~H. and {Busetto}, G. and {Buson}, S. and {Caliandro}, G.~A. and {Cameron}, R.~A. and {Camilo}, F. and {Caraveo}, P.~A. and {Casandjian}, J.~M. and {Cecchi}, C. and {{\c{C}}elik}, {\"O}. and {Charles}, E. and {Chaty}, S. and {Chaves}, R.~C.~G. and {Chekhtman}, A. and {Chen}, A.~W. and {Chiang}, J. and {Chiaro}, G. and {Ciprini}, S. and {Claus}, R. and {Cognard}, I. and {Cohen-Tanugi}, J. and {Cominsky}, L.~R. and {Conrad}, J. and {Cutini}, S. and {D'Ammando}, F. and {de Angelis}, A. and {DeCesar}, M.~E. and {De Luca}, A. and {den Hartog}, P.~R. and {de Palma}, F. and {Dermer}, C.~D. and {Desvignes}, G. and {Digel}, S.~W. and {Di Venere}, L. and {Drell}, P.~S. and {Drlica-Wagner}, A. and {Dubois}, R. and {Dumora}, D. and {Espinoza}, C.~M. and {Falletti}, L. and {Favuzzi}, C. and {Ferrara}, E.~C. and {Focke}, W.~B. and {Franckowiak}, A. and {Freire}, P.~C.~C. and {Funk}, S. and {Fusco}, P. and {Gargano}, F. and {Gasparrini}, D. and {Germani}, S. and {Giglietto}, N. and {Giommi}, P. and {Giordano}, F. and {Giroletti}, M. and {Glanzman}, T. and {Godfrey}, G. and {Gotthelf}, E.~V. and {Grenier}, I.~A. and {Grondin}, M. -H. and {Grove}, J.~E. and {Guillemot}, L. and {Guiriec}, S. and {Hadasch}, D. and {Hanabata}, Y. and {Harding}, A.~K. and {Hayashida}, M. and {Hays}, E. and {Hessels}, J. and {Hewitt}, J. and {Hill}, A.~B. and {Horan}, D. and {Hou}, X. and {Hughes}, R.~E. and {Jackson}, M.~S. and {Janssen}, G.~H. and {Jogler}, T. and {J{\'o}hannesson}, G. and {Johnson}, R.~P. and {Johnson}, A.~S. and {Johnson}, T.~J. and {Johnson}, W.~N. and {Johnston}, S. and {Kamae}, T. and {Kataoka}, J. and {Keith}, M. and {Kerr}, M. and {Kn{\"o}dlseder}, J. and {Kramer}, M. and {Kuss}, M. and {Lande}, J. and {Larsson}, S. and {Latronico}, L. and {Lemoine-Goumard}, M. and {Longo}, F. and {Loparco}, F. and {Lovellette}, M.~N. and {Lubrano}, P. and {Lyne}, A.~G. and {Manchester}, R.~N. and {Marelli}, M. and {Massaro}, F. and {Mayer}, M. and {Mazziotta}, M.~N. and {McEnery}, J.~E. and {McLaughlin}, M.~A. and {Mehault}, J. and {Michelson}, P.~F. and {Mignani}, R.~P. and {Mitthumsiri}, W. and {Mizuno}, T. and {Moiseev}, A.~A. and {Monzani}, M.~E. and {Morselli}, A. and {Moskalenko}, I.~V. and {Murgia}, S. and {Nakamori}, T. and {Nemmen}, R. and {Nuss}, E. and {Ohno}, M. and {Ohsugi}, T. and {Orienti}, M. and {Orlando}, E. and {Ormes}, J.~F. and {Paneque}, D. and {Panetta}, J.~H. and {Parent}, D. and {Perkins}, J.~S. and {Pesce-Rollins}, M. and {Pierbattista}, M. and {Piron}, F. and {Pivato}, G. and {Pletsch}, H.~J. and {Porter}, T.~A. and {Possenti}, A. and {Rain{\`o}}, S. and {Rando}, R. and {Ransom}, S.~M. and {Ray}, P.~S. and {Razzano}, M. and {Rea}, N. and {Reimer}, A. and {Reimer}, O. and {Renault}, N. and {Reposeur}, T. and {Ritz}, S. and {Romani}, R.~W. and {Roth}, M. and {Rousseau}, R. and {Roy}, J. and {Ruan}, J. and {Sartori}, A. and {Saz Parkinson}, P.~M. and {Scargle}, J.~D. and {Schulz}, A. and {Sgr{\`o}}, C. and {Shannon}, R. and {Siskind}, E.~J. and {Smith}, D.~A. and {Spandre}, G. and {Spinelli}, P. and {Stappers}, B.~W. and {Strong}, A.~W. and {Suson}, D.~J. and {Takahashi}, H. and {Thayer}, J.~G. and {Thayer}, J.~B. and {Theureau}, G. and {Thompson}, D.~J. and {Thorsett}, S.~E. and {Tibaldo}, L. and {Tibolla}, O. and {Tinivella}, M. and {Torres}, D.~F. and {Tosti}, G. and {Troja}, E. and {Uchiyama}, Y. and {Usher}, T.~L. and {Vandenbroucke}, J. and {Vasileiou}, V. and {Venter}, C. and {Vianello}, G. and {Vitale}, V. and {Wang}, N. and {Weltevrede}, P. and {Winer}, B.~L. and {Wolff}, M.~T. and {Wood}, D.~L. and {Wood}, K.~S. and {Wood}, M. and {Yang}, Z.},
        title = "{The Second Fermi Large Area Telescope Catalog of Gamma-Ray Pulsars}",
      journal = {\apjs},
         year = 2013,
        month = oct,
       volume = {208},
       number = {2},
          eid = {17},
        pages = {17},
          doi = {10.1088/0067-0049/208/2/17},
archivePrefix = {arXiv},
       eprint = {1305.4385},
 primaryClass = {astro-ph.HE},
       adsurl = {https://ui.adsabs.harvard.edu/abs/2013ApJS..208...17A}
}

@INPROCEEDINGS{Roberts13,
       author = {{Roberts}, Mallory S.~E.},
        title = "{Surrounded by spiders! New black widows and redbacks in the Galactic field}",
    booktitle = {Neutron Stars and Pulsars: Challenges and Opportunities after 80 years},
         year = 2013,
       editor = {{van Leeuwen}, Joeri},
       volume = {291},
        month = mar,
        pages = {127-132},
          doi = {10.1017/S174392131202337X},
archivePrefix = {arXiv},
       eprint = {1210.6903},
 primaryClass = {astro-ph.HE},
       adsurl = {https://ui.adsabs.harvard.edu/abs/2013IAUS..291..127R},
    series = {}
}

@ARTICLE{Strader19,
       author = {{Strader}, Jay and {Swihart}, Samuel and {Chomiuk}, Laura and {Bahramian}, Arash and {Britt}, Chris and {Cheung}, C.~C. and {Dage}, Kristen and {Halpern}, Jules and {Li}, Kwan-Lok and {Mignani}, Roberto P. and {Orosz}, Jerome A. and {Peacock}, Mark and {Salinas}, Ricardo and {Shishkovsky}, Laura and {Tremou}, Evangelia},
        title = "{Optical Spectroscopy and Demographics of Redback Millisecond Pulsar Binaries}",
      journal = {\apj},
         year = 2019,
        month = feb,
       volume = {872},
       number = {1},
          eid = {42},
        pages = {42},
          doi = {10.3847/1538-4357/aafbaa},
archivePrefix = {arXiv},
       eprint = {1812.04626},
 primaryClass = {astro-ph.HE},
       adsurl = {https://ui.adsabs.harvard.edu/abs/2019ApJ...872...42S}
}

@ARTICLE{Kandel20,
       author = {{Kandel}, D. and {Romani}, Roger W.},
        title = "{Atmospheric Circulation on Black Widow Companions}",
      journal = {\apj},
         year = 2020,
        month = apr,
       volume = {892},
       number = {2},
          eid = {101},
        pages = {101},
          doi = {10.3847/1538-4357/ab7b62},
archivePrefix = {arXiv},
       eprint = {2002.12483},
 primaryClass = {astro-ph.HE},
       adsurl = {https://ui.adsabs.harvard.edu/abs/2020ApJ...892..101K}
}

@ARTICLE{Strader15,
       author = {{Strader}, Jay and {Chomiuk}, Laura and {Cheung}, C.~C. and {Sand}, David J. and {Donato}, Davide and {Corbet}, Robin H.~D. and {Koeppe}, Dana and {Edwards}, Philip G. and {Stevens}, Jamie and {Petrov}, Leonid and {Salinas}, Ricardo and {Peacock}, Mark and {Finzell}, Thomas and {Reichart}, Daniel E. and {Haislip}, Joshua B.},
        title = "{1FGL J1417.7-4407: A Likely Gamma-Ray Bright Binary with a Massive Neutron Star and a Giant Secondary}",
      journal = {\apjl},
         year = 2015,
        month = may,
       volume = {804},
       number = {1},
          eid = {L12},
        pages = {L12},
          doi = {10.1088/2041-8205/804/1/L12},
archivePrefix = {arXiv},
       eprint = {1502.05999},
 primaryClass = {astro-ph.HE},
       adsurl = {https://ui.adsabs.harvard.edu/abs/2015ApJ...804L..12S}
}

@ARTICLE{HI4PI,
       author = {{HI4PI Collaboration} and {Ben Bekhti}, N. and {Fl{\"o}er}, L. and {Keller}, R. and {Kerp}, J. and {Lenz}, D. and {Winkel}, B. and {Bailin}, J. and {Calabretta}, M.~R. and {Dedes}, L. and {Ford}, H.~A. and {Gibson}, B.~K. and {Haud}, U. and {Janowiecki}, S. and {Kalberla}, P.~M.~W. and {Lockman}, F.~J. and {McClure-Griffiths}, N.~M. and {Murphy}, T. and {Nakanishi}, H. and {Pisano}, D.~J. and {Staveley-Smith}, L.},
        title = "{HI4PI: A full-sky H I survey based on EBHIS and GASS}",
      journal = {\aap},
         year = 2016,
        month = oct,
       volume = {594},
          eid = {A116},
        pages = {A116},
          doi = {10.1051/0004-6361/201629178},
archivePrefix = {arXiv},
       eprint = {1610.06175},
 primaryClass = {astro-ph.GA},
       adsurl = {https://ui.adsabs.harvard.edu/abs/2016A&A...594A.116H}
}

@INPROCEEDINGS{Tody86,
       author = {{Tody}, Doug},
        title = "{The IRAF Data Reduction and Analysis System}",
    booktitle = {Instrumentation in astronomy VI},
         year = 1986,
       editor = {{Crawford}, David L.},
       series = {Society of Photo-Optical Instrumentation Engineers (SPIE) Conference Series},
       volume = {627},
        month = jan,
        pages = {733},
          doi = {10.1117/12.968154},
       adsurl = {https://ui.adsabs.harvard.edu/abs/1986SPIE..627..733T}
}

@INPROCEEDINGS{Clemens04,
       author = {{Clemens}, J. Christopher and {Crain}, J. Adam and {Anderson}, Robert},
        title = "{The Goodman spectrograph}",
    booktitle = {Ground-based Instrumentation for Astronomy},
         year = 2004,
       editor = {{Moorwood}, Alan F.~M. and {Iye}, Masanori},
       series = {Society of Photo-Optical Instrumentation Engineers (SPIE) Conference Series},
       volume = {5492},
        month = sep,
        pages = {331-340},
          doi = {10.1117/12.550069},
       adsurl = {https://ui.adsabs.harvard.edu/abs/2004SPIE.5492..331C}
}

@ARTICLE{Camilo16,
       author = {{Camilo}, F. and {Reynolds}, J.~E. and {Ransom}, S.~M. and {Halpern}, J.~P. and {Bogdanov}, S. and {Kerr}, M. and {Ray}, P.~S. and {Cordes}, J.~M. and {Sarkissian}, J. and {Barr}, E.~D. and {Ferrara}, E.~C.},
        title = "{Discovery of a Millisecond Pulsar in the 5.4 day Binary 3FGL J1417.5-4402: Observing the Late Phase of Pulsar Recycling}",
      journal = {\apj},
         year = 2016,
        month = mar,
       volume = {820},
       number = {1},
          eid = {6},
        pages = {6},
          doi = {10.3847/0004-637X/820/1/6},
archivePrefix = {arXiv},
       eprint = {1603.00142},
 primaryClass = {astro-ph.SR},
       adsurl = {https://ui.adsabs.harvard.edu/abs/2016ApJ...820....6C}
}

@ARTICLE{Swihart18,
       author = {{Swihart}, Samuel J. and {Strader}, Jay and {Shishkovsky}, Laura and {Chomiuk}, Laura and {Bahramian}, Arash and {Heinke}, Craig O. and {Miller-Jones}, James C.~A. and {Edwards}, Philip G. and {Cheung}, C.~C.},
        title = "{A Multiwavelength View of the Neutron Star Binary 1FGL J1417.7-4402: A Progenitor to Canonical Millisecond Pulsars}",
      journal = {\apj},
         year = 2018,
        month = oct,
       volume = {866},
       number = {2},
          eid = {83},
        pages = {83},
          doi = {10.3847/1538-4357/aadcab},
archivePrefix = {arXiv},
       eprint = {1808.07101},
 primaryClass = {astro-ph.HE},
       adsurl = {https://ui.adsabs.harvard.edu/abs/2018ApJ...866...83S}
}

@ARTICLE{Linares18,
       author = {{Linares}, M. and {Shahbaz}, T. and {Casares}, J.},
        title = "{Peering into the Dark Side: Magnesium Lines Establish a Massive Neutron Star in PSR J2215+5135}",
      journal = {\apj},
         year = 2018,
        month = may,
       volume = {859},
       number = {1},
          eid = {54},
        pages = {54},
          doi = {10.3847/1538-4357/aabde6},
archivePrefix = {arXiv},
       eprint = {1805.08799},
 primaryClass = {astro-ph.HE},
       adsurl = {https://ui.adsabs.harvard.edu/abs/2018ApJ...859...54L}
}

@ARTICLE{Ray12,
       author = {{Ray}, P.~S. and {Abdo}, A.~A. and {Parent}, D. and {Bhattacharya}, D. and {Bhattacharyya}, B. and {Camilo}, F. and {Cognard}, I. and {Theureau}, G. and {Ferrara}, E.~C. and {Harding}, A.~K. and {Thompson}, D.~J. and {Freire}, P.~C.~C. and {Guillemot}, L. and {Gupta}, Y. and {Roy}, J. and {Hessels}, J.~W.~T. and {Johnston}, S. and {Keith}, M. and {Shannon}, R. and {Kerr}, M. and {Michelson}, P.~F. and {Romani}, R.~W. and {Kramer}, M. and {McLaughlin}, M.~A. and {Ransom}, S.~M. and {Roberts}, M.~S.~E. and {Saz Parkinson}, P.~M. and {Ziegler}, M. and {Smith}, D.~A. and {Stappers}, B.~W. and {Weltevrede}, P. and {Wood}, K.~S.},
        title = "{Radio Searches of Fermi LAT Sources and Blind Search Pulsars: The Fermi Pulsar Search Consortium}",
      journal = {arXiv e-prints},
         year = 2012,
        month = may,
          eid = {arXiv:1205.3089},
        pages = {arXiv:1205.3089},
archivePrefix = {arXiv},
       eprint = {1205.3089},
 primaryClass = {astro-ph.HE},
       adsurl = {https://ui.adsabs.harvard.edu/abs/2012arXiv1205.3089R}
}

@ARTICLE{GaiaDR3,
       author = {{Gaia Collaboration} and {Vallenari}, A. and {Brown}, A.~G.~A. and {Prusti}, T. and {de Bruijne}, J.~H.~J. and {Arenou}, F. and {Babusiaux}, C. and {Biermann}, M. and {Creevey}, O.~L. and {Ducourant}, C. and {Evans}, D.~W. and {Eyer}, L. and {Guerra}, R. and {Hutton}, A. and {Jordi}, C. and {Klioner}, S.~A. and {Lammers}, U.~L. and {Lindegren}, L. and {Luri}, X. and {Mignard}, F. and {Panem}, C. and {Pourbaix}, D. and {Randich}, S. and {Sartoretti}, P. and {Soubiran}, C. and {Tanga}, P. and {Walton}, N.~A. and {Bailer-Jones}, C.~A.~L. and {Bastian}, U. and {Drimmel}, R. and {Jansen}, F. and {Katz}, D. and {Lattanzi}, M.~G. and {van Leeuwen}, F. and {Bakker}, J. and {Cacciari}, C. and {Casta{\~n}eda}, J. and {De Angeli}, F. and {Fabricius}, C. and {Fouesneau}, M. and {Fr{\'e}mat}, Y. and {Galluccio}, L. and {Guerrier}, A. and {Heiter}, U. and {Masana}, E. and {Messineo}, R. and {Mowlavi}, N. and {Nicolas}, C. and {Nienartowicz}, K. and {Pailler}, F. and {Panuzzo}, P. and {Riclet}, F. and {Roux}, W. and {Seabroke}, G.~M. and {Sordo{\o}rcit}, R. and {Th{\'e}venin}, F. and {Gracia-Abril}, G. and {Portell}, J. and {Teyssier}, D. and {Altmann}, M. and {Andrae}, R. and {Audard}, M. and {Bellas-Velidis}, I. and {Benson}, K. and {Berthier}, J. and {Blomme}, R. and {Burgess}, P.~W. and {Busonero}, D. and {Busso}, G. and {C{\'a}novas}, H. and {Carry}, B. and {Cellino}, A. and {Cheek}, N. and {Clementini}, G. and {Damerdji}, Y. and {Davidson}, M. and {de Teodoro}, P. and {Nu{\~n}ez Campos}, M. and {Delchambre}, L. and {Dell'Oro}, A. and {Esquej}, P. and {Fern{\'a}ndez-Hern{\'a}ndez}, J. and {Fraile}, E. and {Garabato}, D. and {Garc{\'\i}a-Lario}, P. and {Gosset}, E. and {Haigron}, R. and {Halbwachs}, J. -L. and {Hambly}, N.~C. and {Harrison}, D.~L. and {Hern{\'a}ndez}, J. and {Hestroffer}, D. and {Hodgkin}, S.~T. and {Holl}, B. and {Jan{\ss}en}, K. and {Jevardat de Fombelle}, G. and {Jordan}, S. and {Krone-Martins}, A. and {Lanzafame}, A.~C. and {L{\"o}ffler}, W. and {Marchal}, O. and {Marrese}, P.~M. and {Moitinho}, A. and {Muinonen}, K. and {Osborne}, P. and {Pancino}, E. and {Pauwels}, T. and {Recio-Blanco}, A. and {Reyl{\'e}}, C. and {Riello}, M. and {Rimoldini}, L. and {Roegiers}, T. and {Rybizki}, J. and {Sarro}, L.~M. and {Siopis}, C. and {Smith}, M. and {Sozzetti}, A. and {Utrilla}, E. and {van Leeuwen}, M. and {Abbas}, U. and {{\'A}brah{\'a}m}, P. and {Abreu Aramburu}, A. and {Aerts}, C. and {Aguado}, J.~J. and {Ajaj}, M. and {Aldea-Montero}, F. and {Altavilla}, G. and {{\'A}lvarez}, M.~A. and {Alves}, J. and {Anders}, F. and {Anderson}, R.~I. and {Anglada Varela}, E. and {Antoja}, T. and {Baines}, D. and {Baker}, S.~G. and {Balaguer-N{\'u}{\~n}ez}, L. and {Balbinot}, E. and {Balog}, Z. and {Barache}, C. and {Barbato}, D. and {Barros}, M. and {Barstow}, M.~A. and {Bartolom{\'e}}, S. and {Bassilana}, J. -L. and {Bauchet}, N. and {Becciani}, U. and {Bellazzini}, M. and {Berihuete}, A. and {Bernet}, M. and {Bertone}, S. and {Bianchi}, L. and {Binnenfeld}, A. and {Blanco-Cuaresma}, S. and {Blazere}, A. and {Boch}, T. and {Bombrun}, A. and {Bossini}, D. and {Bouquillon}, S. and {Bragaglia}, A. and {Bramante}, L. and {Breedt}, E. and {Bressan}, A. and {Brouillet}, N. and {Brugaletta}, E. and {Bucciarelli}, B. and {Burlacu}, A. and {Butkevich}, A.~G. and {Buzzi}, R. and {Caffau}, E. and {Cancelliere}, R. and {Cantat-Gaudin}, T. and {Carballo}, R. and {Carlucci}, T. and {Carnerero}, M.~I. and {Carrasco}, J.~M. and {Casamiquela}, L. and {Castellani}, M. and {Castro-Ginard}, A. and {Chaoul}, L. and {Charlot}, P. and {Chemin}, L. and {Chiaramida}, V. and {Chiavassa}, A. and {Chornay}, N. and {Comoretto}, G. and {Contursi}, G. and {Cooper}, W.~J. and {Cornez}, T. and {Cowell}, S. and {Crifo}, F. and {Cropper}, M. and {Crosta}, M. and {Crowley}, C. and {Dafonte}, C. and {Dapergolas}, A. and {David}, M. and {David}, P. and {de Laverny}, P. and {De Luise}, F. and {De March}, R. and {De Ridder}, J. and {de Souza}, R. and {de Torres}, A. and {del Peloso}, E.~F. and {del Pozo}, E. and {Delbo}, M. and {Delgado}, A. and {Delisle}, J. -B. and {Demouchy}, C. and {Dharmawardena}, T.~E. and {Di Matteo}, P. and {Diakite}, S. and {Diener}, C. and {Distefano}, E. and {Dolding}, C. and {Edvardsson}, B. and {Enke}, H. and {Fabre}, C. and {Fabrizio}, M. and {Faigler}, S. and {Fedorets}, G. and {Fernique}, P. and {Fienga}, A. and {Figueras}, F. and {Fournier}, Y. and {Fouron}, C. and {Fragkoudi}, F. and {Gai}, M. and {Garcia-Gutierrez}, A. and {Garcia-Reinaldos}, M. and {Garc{\'\i}a-Torres}, M. and {Garofalo}, A. and {Gavel}, A. and {Gavras}, P. and {Gerlach}, E. and {Geyer}, R. and {Giacobbe}, P. and {Gilmore}, G. and {Girona}, S. and {Giuffrida}, G. and {Gomel}, R. and {Gomez}, A. and {Gonz{\'a}lez-N{\'u}{\~n}ez}, J. and {Gonz{\'a}lez-Santamar{\'\i}a}, I. and {Gonz{\'a}lez-Vidal}, J.~J. and {Granvik}, M. and {Guillout}, P. and {Guiraud}, J. and {Guti{\'e}rrez-S{\'a}nchez}, R. and {Guy}, L.~P. and {Hatzidimitriou}, D. and {Hauser}, M. and {Haywood}, M. and {Helmer}, A. and {Helmi}, A. and {Sarmiento}, M.~H. and {Hidalgo}, S.~L. and {Hilger}, T. and {H{\l}adczuk}, N. and {Hobbs}, D. and {Holland}, G. and {Huckle}, H.~E. and {Jardine}, K. and {Jasniewicz}, G. and {Jean-Antoine Piccolo}, A. and {Jim{\'e}nez-Arranz}, {\'O}. and {Jorissen}, A. and {Juaristi Campillo}, J. and {Julbe}, F. and {Karbevska}, L. and {Kervella}, P. and {Khanna}, S. and {Kontizas}, M. and {Kordopatis}, G. and {Korn}, A.~J. and {K{\'o}sp{\'a}l}, {\'A} and {Kostrzewa-Rutkowska}, Z. and {Kruszy{\'n}ska}, K. and {Kun}, M. and {Laizeau}, P. and {Lambert}, S. and {Lanza}, A.~F. and {Lasne}, Y. and {Le Campion}, J. -F. and {Lebreton}, Y. and {Lebzelter}, T. and {Leccia}, S. and {Leclerc}, N. and {Lecoeur-Taibi}, I. and {Liao}, S. and {Licata}, E.~L. and {Lindstr{\o}m}, H.~E.~P. and {Lister}, T.~A. and {Livanou}, E. and {Lobel}, A. and {Lorca}, A. and {Loup}, C. and {Madrero Pardo}, P. and {Magdaleno Romeo}, A. and {Managau}, S. and {Mann}, R.~G. and {Manteiga}, M. and {Marchant}, J.~M.},
        title = "{Gaia Data Release 3: Summary of the content and survey properties}",
      journal = {arXiv e-prints},
         year = 2022,
        month = jul,
          eid = {arXiv:2208.00211},
        pages = {arXiv:2208.00211},
archivePrefix = {arXiv},
       eprint = {2208.00211},
 primaryClass = {astro-ph.GA},
       adsurl = {https://ui.adsabs.harvard.edu/abs/2022arXiv220800211G}
}

@ARTICLE{Sabbi2003,
       author = {{Sabbi}, E. and {Gratton}, R. and {Ferraro}, F.~R. and {Bragaglia}, A. and {Possenti}, A. and {D'Amico}, N. and {Camilo}, F.},
        title = "{The Complex H{\ensuremath{\alpha}} Line Profile of the Bright Companion to PSR J1740-5340 in NGC 6397}",
      journal = {The Astrophysical Journal},
         year = 2003,
        month = may,
       volume = {589},
       number = {1},
        pages = {L41-L44},
          doi = {10.1086/375729},
archivePrefix = {arXiv},
       eprint = {astro-ph/0304159},
 primaryClass = {astro-ph},
       adsurl = {https://ui.adsabs.harvard.edu/abs/2003ApJ...589L..41S}
}

@ARTICLE{Lindegren2021,
       author = {{Lindegren}, L. and {Bastian}, U. and {Biermann}, M. and {Bombrun}, A. and {de Torres}, A. and {Gerlach}, E. and {Geyer}, R. and {Hern{\'a}ndez}, J. and {Hilger}, T. and {Hobbs}, D. and {Klioner}, S.~A. and {Lammers}, U. and {McMillan}, P.~J. and {Ramos-Lerate}, M. and {Steidelm{\"u}ller}, H. and {Stephenson}, C.~A. and {van Leeuwen}, F.},
        title = "{Gaia Early Data Release 3. Parallax bias versus magnitude, colour, and position}",
      journal = {Astronomy and Astrophysics},
         year = 2021,
        month = may,
       volume = {649},
          eid = {A4},
        pages = {A4},
          doi = {10.1051/0004-6361/202039653},
archivePrefix = {arXiv},
       eprint = {2012.01742},
 primaryClass = {astro-ph.IM},
       adsurl = {https://ui.adsabs.harvard.edu/abs/2021A&A...649A...4L}
}

@ARTICLE{Evans2023,
       author = {{Evans}, P.~A. and {Page}, K.~L. and {Beardmore}, A.~P. and {Eyles-Ferris}, R.~A.~J. and {Osborne}, J.~P. and {Campana}, S. and {Kennea}, J.~A. and {Cenko}, S.~B.},
        title = "{A real-time transient detector and the living Swift-XRT point source catalogue}",
      journal = {Monthly Notices of the Royal Astronomical Society},
         year = 2023,
        month = jan,
       volume = {518},
       number = {1},
        pages = {174-184},
          doi = {10.1093/mnras/stac2937},
archivePrefix = {arXiv},
       eprint = {2208.14478},
 primaryClass = {astro-ph.HE},
       adsurl = {https://ui.adsabs.harvard.edu/abs/2023MNRAS.518..174E}
}

@INPROCEEDINGS{Tody93,
       author = {{Tody}, Doug},
        title = "{IRAF in the Nineties}",
    booktitle = {Astronomical Data Analysis Software and Systems II},
         year = 1993,
       editor = {{Hanisch}, R.~J. and {Brissenden}, R.~J.~V. and {Barnes}, J.},
       series = {Astronomical Society of the Pacific Conference Series},
       volume = {52},
        month = jan,
        pages = {173},
       adsurl = {https://ui.adsabs.harvard.edu/abs/1993ASPC...52..173T}
}

@ARTICLE{Fitzpatrick24,
       author = {{Fitzpatrick}, Michael and {Placco}, Vinicius and {Bolton}, Adam and {Merino}, Brian and {Ridgway}, Susan and {Stanghellini}, Letizia},
        title = "{Modernizing IRAF to Support Gemini Data Reduction}",
      journal = {arXiv e-prints},
         year = 2024,
        month = jan,
          eid = {arXiv:2401.01982},
        pages = {arXiv:2401.01982},
          doi = {10.48550/arXiv.2401.01982},
archivePrefix = {arXiv},
       eprint = {2401.01982},
 primaryClass = {astro-ph.IM},
       adsurl = {https://ui.adsabs.harvard.edu/abs/2024arXiv240101982F}
}

@ARTICLE{Gullikson2014,
       author = {{Gullikson}, Kevin and {Dodson-Robinson}, Sarah and {Kraus}, Adam},
        title = "{Correcting for Telluric Absorption: Methods, Case Studies, and Release of the TelFit Code}",
      journal = {The Astronomical Journal},
         year = 2014,
        month = sep,
       volume = {148},
       number = {3},
          eid = {53},
        pages = {53},
          doi = {10.1088/0004-6256/148/3/53},
archivePrefix = {arXiv},
       eprint = {1406.6059},
 primaryClass = {astro-ph.IM},
       adsurl = {https://ui.adsabs.harvard.edu/abs/2014AJ....148...53G}
}

@ARTICLE{Dodge2024,
       author = {{Dodge}, O.~G. and {Breton}, R.~P. and {Clark}, C.~J. and {Burgay}, M. and {Strader}, J. and {Au}, K. -Y. and {Barr}, E.~D. and {Buchner}, S. and {Dhillon}, V.~S. and {Ferrara}, E.~C. and {Freire}, P.~C.~C. and {Griessmeier}, J. -M. and {Kennedy}, M.~R. and {Kramer}, M. and {Li}, K. -L. and {Padmanabh}, P.~V. and {Phosrisom}, A. and {Stappers}, B.~W. and {Swihart}, S.~J. and {Thongmeearkom}, T.},
        title = "{Mass estimates from optical modelling of the new TRAPUM redback PSR J1910-5320}",
      journal = {Monthly Notices of the Royal Astronomical Society},
         year = 2024,
        month = mar,
       volume = {528},
       number = {3},
        pages = {4337-4353},
          doi = {10.1093/mnras/stae211},
archivePrefix = {arXiv},
       eprint = {2401.09928},
 primaryClass = {astro-ph.HE},
       adsurl = {https://ui.adsabs.harvard.edu/abs/2024MNRAS.528.4337D}
}

@ARTICLE{Onken2024,
       author = {{Onken}, Christopher A. and {Wolf}, Christian and {Bessell}, Michael S. and {Chang}, Seo-Won and {Luvaul}, Lance C. and {Tonry}, John L. and {White}, Marc C. and {Da Costa}, Gary S.},
        title = "{SkyMapper Southern Survey: Data Release 4}",
      journal = {arXiv e-prints},
         year = 2024,
        month = feb,
          eid = {arXiv:2402.02015},
        pages = {arXiv:2402.02015},
          doi = {10.48550/arXiv.2402.02015},
archivePrefix = {arXiv},
       eprint = {2402.02015},
 primaryClass = {astro-ph.CO},
       adsurl = {https://ui.adsabs.harvard.edu/abs/2024arXiv240202015O}
}

@ARTICLE{Koposov2011,
       author = {{Koposov}, Sergey E. and {Gilmore}, G. and {Walker}, M.~G. and {Belokurov}, V. and {Evans}, N. Wyn and {Fellhauer}, M. and {Gieren}, W. and {Geisler}, D. and {Monaco}, L. and {Norris}, J.~E. and {Okamoto}, S. and {Pe{\~n}arrubia}, J. and {Wilkinson}, M. and {Wyse}, R.~F.~G. and {Zucker}, D.~B.},
        title = "{Accurate Stellar Kinematics at Faint Magnitudes: Application to the Bo{\"o}tes I Dwarf Spheroidal Galaxy}",
      journal = {The Astrophysical Journal},
         year = 2011,
        month = aug,
       volume = {736},
       number = {2},
          eid = {146},
        pages = {146},
          doi = {10.1088/0004-637X/736/2/146},
archivePrefix = {arXiv},
       eprint = {1105.4102},
 primaryClass = {astro-ph.GA},
       adsurl = {https://ui.adsabs.harvard.edu/abs/2011ApJ...736..146K}
}

@software{Koposov2019,
       author = {{Koposov}, Sergey E.},
        title = "{RVSpecFit: Radial velocity and stellar atmospheric parameter fitting}",
 howpublished = {Astrophysics Source Code Library, record ascl:1907.013},
         year = 2019,
        month = jul,
          eid = {ascl:1907.013},
       adsurl = {https://ui.adsabs.harvard.edu/abs/2019ascl.soft07013K}
}

@INPROCEEDINGS{Allard2016,
       author = {{Allard}, F.},
        title = "{The PHOENIX Model Atmosphere Grid for Stars}",
    booktitle = {SF2A-2016: Proceedings of the Annual meeting of the French Society of Astronomy and Astrophysics},
         year = 2016,
       editor = {{Reyl{\'e}}, C. and {Richard}, J. and {Cambr{\'e}sy}, L. and {Deleuil}, M. and {P{\'e}contal}, E. and {Tresse}, L. and {Vauglin}, I.},
        month = dec,
        pages = {223-227},
       adsurl = {https://ui.adsabs.harvard.edu/abs/2016sf2a.conf..223A}
}

@ARTICLE{Price-Whelan2017,
       author = {{Price-Whelan}, Adrian M. and {Hogg}, David W. and {Foreman-Mackey}, Daniel and {Rix}, Hans-Walter},
        title = "{The Joker: A Custom Monte Carlo Sampler for Binary-star and Exoplanet Radial Velocity Data}",
      journal = {The Astrophysical Journal},
         year = 2017,
        month = mar,
       volume = {837},
       number = {1},
          eid = {20},
        pages = {20},
          doi = {10.3847/1538-4357/aa5e50},
archivePrefix = {arXiv},
       eprint = {1610.07602},
 primaryClass = {astro-ph.SR},
       adsurl = {https://ui.adsabs.harvard.edu/abs/2017ApJ...837...20P}
}

@ARTICLE{Tonry2018,
       author = {{Tonry}, J.~L. and {Denneau}, L. and {Heinze}, A.~N. and {Stalder}, B. and {Smith}, K.~W. and {Smartt}, S.~J. and {Stubbs}, C.~W. and {Weiland}, H.~J. and {Rest}, A.},
        title = "{ATLAS: A High-cadence All-sky Survey System}",
      journal = {Publications of the Astronomical Society of the Pacific},
         year = 2018,
        month = jun,
       volume = {130},
       number = {988},
        pages = {064505},
          doi = {10.1088/1538-3873/aabadf},
archivePrefix = {arXiv},
       eprint = {1802.00879},
 primaryClass = {astro-ph.IM},
       adsurl = {https://ui.adsabs.harvard.edu/abs/2018PASP..130f4505T}
}

@ARTICLE{Heinze2018,
       author = {{Heinze}, A.~N. and {Tonry}, J.~L. and {Denneau}, L. and {Flewelling}, H. and {Stalder}, B. and {Rest}, A. and {Smith}, K.~W. and {Smartt}, S.~J. and {Weiland}, H.},
        title = "{A First Catalog of Variable Stars Measured by the Asteroid Terrestrial-impact Last Alert System (ATLAS)}",
      journal = {The Astronomical Journal},
         year = 2018,
        month = nov,
       volume = {156},
       number = {5},
          eid = {241},
        pages = {241},
          doi = {10.3847/1538-3881/aae47f},
archivePrefix = {arXiv},
       eprint = {1804.02132},
 primaryClass = {astro-ph.SR},
       adsurl = {https://ui.adsabs.harvard.edu/abs/2018AJ....156..241H}
}

@ARTICLE{Shingles2021,
       author = {{Shingles}, L. and {Smith}, K.~W. and {Young}, D.~R. and {Smartt}, S.~J. and {Tonry}, J. and {Denneau}, L. and {Heinze}, A. and {Weiland}, H. and {Flewelling}, H. and {Stalder}, B. and {Clocchiatti}, A. and {F{\"o}rster}, F. and {Pignata}, G. and {Rest}, A. and {Anderson}, J. and {Stubbs}, C. and {Erasmus}, N.},
        title = "{Release of the ATLAS Forced Photometry server for public use}",
      journal = {Transient Name Server AstroNote},
         year = 2021,
        month = jan,
       volume = {7},
        pages = {1-7},
       adsurl = {https://ui.adsabs.harvard.edu/abs/2021TNSAN...7....1S}
}

@ARTICLE{Davey1992,
       author = {{Davey}, Stephen and {Smith}, Robert C.},
        title = "{Irradiation of the secondary star in cataclysmic variables.}",
      journal = {Monthly Notices of the Royal Astronomical Society},
         year = 1992,
        month = aug,
       volume = {257},
        pages = {476-484},
          doi = {10.1093/mnras/257.3.476},
       adsurl = {https://ui.adsabs.harvard.edu/abs/1992MNRAS.257..476D}
}

@ARTICLE{Turchetta2023,
       author = {{Turchetta}, Marco and {Linares}, Manuel and {Koljonen}, Karri and {Sen}, Bidisha},
        title = "{Quantifying irradiation in spider pulsars: the extreme case of PSR J1622-0315}",
      journal = {Monthly Notices of the Royal Astronomical Society},
         year = 2023,
        month = oct,
       volume = {525},
       number = {2},
        pages = {2565-2570},
          doi = {10.1093/mnras/stad2435},
archivePrefix = {arXiv},
       eprint = {2307.06330},
 primaryClass = {astro-ph.HE},
       adsurl = {https://ui.adsabs.harvard.edu/abs/2023MNRAS.525.2565T}
}

@ARTICLE{Karpova2026,
       author = {{Karpova}, A.~V. and {Zyuzin}, D.~A. and {Zharikov}, S.~V. and {Gilfanov}, M.~R.},
        title = "{Nature of 4FGL J2249.4+6229: evidence for a redback system with a cool companion and low X-ray and {\ensuremath{\gamma}}-ray luminosities}",
      journal = {Monthly Notices of the Royal Astronomical Society},
         year = 2026,
        month = jul,
       volume = {549},
       number = {3},
          eid = {stag971},
        pages = {stag971},
          doi = {10.1093/mnras/stag971},
archivePrefix = {arXiv},
       eprint = {2605.22271},
 primaryClass = {astro-ph.HE},
       adsurl = {https://ui.adsabs.harvard.edu/abs/2026MNRAS.549ag971K}
}

@ARTICLE{Ballet2026,
       author = {{Ballet}, J. and {Bruel}, P. and {Burnett}, T.~H. and {Lott}, B.},
        title = "{Fermi-LAT 16-year Source List}",
      journal = {arXiv e-prints},
         year = 2026,
        month = feb,
          eid = {arXiv:2602.22148},
        pages = {arXiv:2602.22148},
          doi = {10.48550/arXiv.2602.22148},
archivePrefix = {arXiv},
       eprint = {2602.22148},
 primaryClass = {astro-ph.HE},
       adsurl = {https://ui.adsabs.harvard.edu/abs/2026arXiv260222148B}
}

@ARTICLE{Simpson2025,
       author = {{Simpson}, Jordan A. and {Linares}, Manuel and {Casares}, Jorge and {Shahbaz}, Tariq and {Sen}, Bidisha and {Camilo}, Fernando},
        title = "{A GTC spectroscopic study of three spider pulsar companions: line-based temperatures, a new face-on redback, and improved mass constraints}",
      journal = {Monthly Notices of the Royal Astronomical Society},
         year = 2025,
        month = jan,
       volume = {536},
       number = {3},
        pages = {2169-2186},
          doi = {10.1093/mnras/stae2728},
archivePrefix = {arXiv},
       eprint = {2408.11099},
 primaryClass = {astro-ph.HE},
       adsurl = {https://ui.adsabs.harvard.edu/abs/2025MNRAS.536.2169S}
}

@ARTICLE{Romani2015,
       author = {{Romani}, Roger W. and {Filippenko}, Alexei V. and {Cenko}, S. Bradley},
        title = "{A Spectroscopic Study of the Extreme Black Widow PSR J1311-3430}",
      journal = {The Astrophysical Journal},
         year = 2015,
        month = may,
       volume = {804},
       number = {2},
          eid = {115},
        pages = {115},
          doi = {10.1088/0004-637X/804/2/115},
archivePrefix = {arXiv},
       eprint = {1503.05247},
 primaryClass = {astro-ph.HE},
       adsurl = {https://ui.adsabs.harvard.edu/abs/2015ApJ...804..115R}
}

@ARTICLE{Swihart2022,
       author = {{Swihart}, Samuel J. and {Strader}, Jay and {Chomiuk}, Laura and {Aydi}, Elias and {Sokolovsky}, Kirill V. and {Ray}, Paul S. and {Kerr}, Matthew},
        title = "{A New Flaring Black Widow Candidate and Demographics of Black Widow Millisecond Pulsars in the Galactic Field}",
      journal = {The Astrophysical Journal},
         year = 2022,
        month = dec,
       volume = {941},
       number = {2},
          eid = {199},
        pages = {199},
          doi = {10.3847/1538-4357/aca2ac},
archivePrefix = {arXiv},
       eprint = {2210.16295},
 primaryClass = {astro-ph.HE},
       adsurl = {https://ui.adsabs.harvard.edu/abs/2022ApJ...941..199S}
}

@ARTICLE{Wang2020,
       author = {{Wang}, Zhongxiang and {Xing}, Yi and {Zhang}, Jujia and {Boutsia}, Konstantina and {Wang}, Gege and {V}, Jithesh and {Burdge}, Kevin B. and {Coughlin}, Michael W. and {Duev}, Dmitry A. and {Kulkarni}, S.~R. and {Riddle}, Reed and {Serabyn}, Eugene},
        title = "{A compact X-ray emitting binary in likely association with 4FGL J0935.3+0901}",
      journal = {Monthly Notices of the Royal Astronomical Society},
         year = 2020,
        month = apr,
       volume = {493},
       number = {4},
        pages = {4845-4851},
          doi = {10.1093/mnras/staa655},
archivePrefix = {arXiv},
       eprint = {2003.08509},
 primaryClass = {astro-ph.HE},
       adsurl = {https://ui.adsabs.harvard.edu/abs/2020MNRAS.493.4845W}
}

@ARTICLE{Mucciarelli2013,
       author = {{Mucciarelli}, A. and {Salaris}, M. and {Lanzoni}, B. and {Pallanca}, C. and {Dalessandro}, E. and {Ferraro}, F.~R.},
        title = "{New Clues on the Nature of the Companion to PSR J1740-5340 in NGC 6397 from XSHOOTER Spectroscopy}",
      journal = {The Astrophysical Journal},
         year = 2013,
        month = aug,
       volume = {772},
       number = {2},
          eid = {L27},
        pages = {L27},
          doi = {10.1088/2041-8205/772/2/L27},
archivePrefix = {arXiv},
       eprint = {1307.0919},
 primaryClass = {astro-ph.SR},
       adsurl = {https://ui.adsabs.harvard.edu/abs/2013ApJ...772L..27M}
}

@ARTICLE{Bogdanov2010,
       author = {{Bogdanov}, Slavko and {van den Berg}, Maureen and {Heinke}, Craig O. and {Cohn}, Haldan N. and {Lugger}, Phyllis M. and {Grindlay}, Jonathan E.},
        title = "{A Chandra X-ray Observatory Study of PSR J1740-5340 and Candidate Millisecond Pulsars in the Globular Cluster NGC 6397}",
      journal = {The Astrophysical Journal},
         year = 2010,
        month = jan,
       volume = {709},
       number = {1},
        pages = {241-250},
          doi = {10.1088/0004-637X/709/1/241},
archivePrefix = {arXiv},
       eprint = {0911.3146},
 primaryClass = {astro-ph.HE},
       adsurl = {https://ui.adsabs.harvard.edu/abs/2010ApJ...709..241B}
}

@ARTICLE{Wadiasingh2017,
       author = {{Wadiasingh}, Zorawar and {Harding}, Alice K. and {Venter}, Christo and {B{\"o}ttcher}, Markus and {Baring}, Matthew G.},
        title = "{Constraining Relativistic Bow Shock Properties in Rotation-powered Millisecond Pulsar Binaries}",
      journal = {The Astrophysical Journal},
         year = 2017,
        month = apr,
       volume = {839},
       number = {2},
          eid = {80},
        pages = {80},
          doi = {10.3847/1538-4357/aa69bf},
archivePrefix = {arXiv},
       eprint = {1703.09560},
 primaryClass = {astro-ph.HE},
       adsurl = {https://ui.adsabs.harvard.edu/abs/2017ApJ...839...80W}
}

@ARTICLE{Kandel2019,
       author = {{Kandel}, D. and {Romani}, Roger W. and {An}, Hongjun},
        title = "{The Synchrotron Emission Pattern of Intrabinary Shocks}",
      journal = {The Astrophysical Journal},
         year = 2019,
        month = jul,
       volume = {879},
       number = {2},
          eid = {73},
        pages = {73},
          doi = {10.3847/1538-4357/ab24d9},
archivePrefix = {arXiv},
       eprint = {1905.12591},
 primaryClass = {astro-ph.HE},
       adsurl = {https://ui.adsabs.harvard.edu/abs/2019ApJ...879...73K}
}

@ARTICLE{Urquhart2020,
       author = {{Urquhart}, Ryan and {Bahramian}, Arash and {Strader}, Jay and {Chomiuk}, Laura and {Ransom}, Scott M. and {Wang}, Yuankun and {Heinke}, Craig and {Tudor}, Vlad and {Miller-Jones}, James C.~A. and {Tetarenko}, Alexandra J. and {Maccarone}, Thomas J. and {Sivakoff}, Gregory R. and {Shishkovsky}, Laura and {Swihart}, Samuel J. and {Tremou}, Evangelia},
        title = "{The MAVERIC Survey: New Compact Binaries Revealed by Deep Radio Continuum Observations of the Galactic Globular Cluster Terzan 5}",
      journal = {The Astrophysical Journal},
         year = 2020,
        month = dec,
       volume = {904},
       number = {2},
          eid = {147},
        pages = {147},
          doi = {10.3847/1538-4357/abb6fc},
archivePrefix = {arXiv},
       eprint = {2009.07286},
 primaryClass = {astro-ph.HE},
       adsurl = {https://ui.adsabs.harvard.edu/abs/2020ApJ...904..147U}
}

@ARTICLE{Koljonen2023,
       author = {{Koljonen}, Karri I.~I. and {Linares}, Manuel},
        title = "{A Gaia view of the optical and X-ray luminosities of compact binary millisecond pulsars}",
      journal = {Monthly Notices of the Royal Astronomical Society},
         year = 2023,
        month = nov,
       volume = {525},
       number = {3},
        pages = {3963-3985},
          doi = {10.1093/mnras/stad2485},
archivePrefix = {arXiv},
       eprint = {2308.07377},
 primaryClass = {astro-ph.HE},
       adsurl = {https://ui.adsabs.harvard.edu/abs/2023MNRAS.525.3963K}
}

@ARTICLE{Miller2020,
       author = {{Miller}, Jessie M. and {Swihart}, Samuel J. and {Strader}, Jay and {Urquhart}, Ryan and {Aydi}, Elias and {Chomiuk}, Laura and {Dage}, Kristen C. and {Kawash}, Adam and {Shishkovsky}, Laura and {Sokolovsky}, Kirill V.},
        title = "{A New Candidate Transitional Millisecond Pulsar in the Subluminous Disk State: 4FGL J0407.7-5702}",
      journal = {The Astrophysical Journal},
         year = 2020,
        month = nov,
       volume = {904},
       number = {1},
          eid = {49},
        pages = {49},
          doi = {10.3847/1538-4357/abbb2e},
archivePrefix = {arXiv},
       eprint = {2009.09054},
 primaryClass = {astro-ph.HE},
       adsurl = {https://ui.adsabs.harvard.edu/abs/2020ApJ...904...49M}
}

@ARTICLE{Zucker2025,
       author = {{Zucker}, Catherine and {Saydjari}, Andrew K. and {Speagle}, Joshua S. and {Schlafly}, Edward F. and {Green}, Gregory M. and {Benjamin}, Robert and {Peek}, Joshua and {Edenhofer}, Gordian and {Goodman}, Alyssa and {Kuhn}, Michael A. and {Finkbeiner}, Douglas P.},
        title = "{A Deep, High-angular-resolution 3D Dust Map of the Southern Galactic Plane}",
      journal = {The Astrophysical Journal},
         year = 2025,
        month = oct,
       volume = {992},
       number = {1},
          eid = {39},
        pages = {39},
          doi = {10.3847/1538-4357/adfbe6},
archivePrefix = {arXiv},
       eprint = {2503.02657},
 primaryClass = {astro-ph.GA},
       adsurl = {https://ui.adsabs.harvard.edu/abs/2025ApJ...992...39Z}
}

@ARTICLE{Choi2016,
       author = {{Choi}, Jieun and {Dotter}, Aaron and {Conroy}, Charlie and {Cantiello}, Matteo and {Paxton}, Bill and {Johnson}, Benjamin D.},
        title = "{Mesa Isochrones and Stellar Tracks (MIST). I. Solar-scaled Models}",
      journal = {The Astrophysical Journal},
         year = 2016,
        month = jun,
       volume = {823},
       number = {2},
          eid = {102},
        pages = {102},
          doi = {10.3847/0004-637X/823/2/102},
archivePrefix = {arXiv},
       eprint = {1604.08592},
 primaryClass = {astro-ph.SR},
       adsurl = {https://ui.adsabs.harvard.edu/abs/2016ApJ...823..102C}
}

@ARTICLE{Dotter2016,
       author = {{Dotter}, Aaron},
        title = "{MESA Isochrones and Stellar Tracks (MIST) 0: Methods for the Construction of Stellar Isochrones}",
      journal = {The Astrophysical Journal Supplement Series},
         year = 2016,
        month = jan,
       volume = {222},
       number = {1},
          eid = {8},
        pages = {8},
          doi = {10.3847/0067-0049/222/1/8},
archivePrefix = {arXiv},
       eprint = {1601.05144},
 primaryClass = {astro-ph.SR},
       adsurl = {https://ui.adsabs.harvard.edu/abs/2016ApJS..222....8D}
}

@ARTICLE{Schlafly2011,
       author = {{Schlafly}, Edward F. and {Finkbeiner}, Douglas P.},
        title = "{Measuring Reddening with Sloan Digital Sky Survey Stellar Spectra and Recalibrating SFD}",
      journal = {The Astrophysical Journal},
         year = 2011,
        month = aug,
       volume = {737},
       number = {2},
          eid = {103},
        pages = {103},
          doi = {10.1088/0004-637X/737/2/103},
archivePrefix = {arXiv},
       eprint = {1012.4804},
 primaryClass = {astro-ph.GA},
       adsurl = {https://ui.adsabs.harvard.edu/abs/2011ApJ...737..103S}
}

@ARTICLE{Abdollahi2020,
       author = {{Abdollahi}, S. and {Acero}, F. and {Ackermann}, M. and {Ajello}, M. and {Atwood}, W.~B. and {Axelsson}, M. and {Baldini}, L. and {Ballet}, J. and {Barbiellini}, G. and {Bastieri}, D. and {Becerra Gonzalez}, J. and {Bellazzini}, R. and {Berretta}, A. and {Bissaldi}, E. and {Blandford}, R.~D. and {Bloom}, E.~D. and {Bonino}, R. and {Bottacini}, E. and {Brandt}, T.~J. and {Bregeon}, J. and {Bruel}, P. and {Buehler}, R. and {Burnett}, T.~H. and {Buson}, S. and {Cameron}, R.~A. and {Caputo}, R. and {Caraveo}, P.~A. and {Casandjian}, J.~M. and {Castro}, D. and {Cavazzuti}, E. and {Charles}, E. and {Chaty}, S. and {Chen}, S. and {Cheung}, C.~C. and {Chiaro}, G. and {Ciprini}, S. and {Cohen-Tanugi}, J. and {Cominsky}, L.~R. and {Coronado-Bl{\'a}zquez}, J. and {Costantin}, D. and {Cuoco}, A. and {Cutini}, S. and {D'Ammando}, F. and {DeKlotz}, M. and {de la Torre Luque}, P. and {de Palma}, F. and {Desai}, A. and {Digel}, S.~W. and {Di Lalla}, N. and {Di Mauro}, M. and {Di Venere}, L. and {Dom{\'\i}nguez}, A. and {Dumora}, D. and {Fana Dirirsa}, F. and {Fegan}, S.~J. and {Ferrara}, E.~C. and {Franckowiak}, A. and {Fukazawa}, Y. and {Funk}, S. and {Fusco}, P. and {Gargano}, F. and {Gasparrini}, D. and {Giglietto}, N. and {Giommi}, P. and {Giordano}, F. and {Giroletti}, M. and {Glanzman}, T. and {Green}, D. and {Grenier}, I.~A. and {Griffin}, S. and {Grondin}, M.-H. and {Grove}, J.~E. and {Guiriec}, S. and {Harding}, A.~K. and {Hayashi}, K. and {Hays}, E. and {Hewitt}, J.~W. and {Horan}, D. and {J{\'o}hannesson}, G. and {Johnson}, T.~J. and {Kamae}, T. and {Kerr}, M. and {Kocevski}, D. and {Kovac'evic'}, M. and {Kuss}, M. and {Landriu}, D. and {Larsson}, S. and {Latronico}, L. and {Lemoine-Goumard}, M. and {Li}, J. and {Liodakis}, I. and {Longo}, F. and {Loparco}, F. and {Lott}, B. and {Lovellette}, M.~N. and {Lubrano}, P. and {Madejski}, G.~M. and {Maldera}, S. and {Malyshev}, D. and {Manfreda}, A. and {Marchesini}, E.~J. and {Marcotulli}, L. and {Mart{\'\i}-Devesa}, G. and {Martin}, P. and {Massaro}, F. and {Mazziotta}, M.~N. and {McEnery}, J.~E. and {Mereu}, I. and {Meyer}, M. and {Michelson}, P.~F. and {Mirabal}, N. and {Mizuno}, T. and {Monzani}, M.~E. and {Morselli}, A. and {Moskalenko}, I.~V. and {Negro}, M. and {Nuss}, E. and {Ojha}, R. and {Omodei}, N. and {Orienti}, M. and {Orlando}, E. and {Ormes}, J.~F. and {Palatiello}, M. and {Paliya}, V.~S. and {Paneque}, D. and {Pei}, Z. and {Pe{\~n}a-Herazo}, H. and {Perkins}, J.~S. and {Persic}, M. and {Pesce-Rollins}, M. and {Petrosian}, V. and {Petrov}, L. and {Piron}, F. and {Poon}, H. and {Porter}, T.~A. and {Principe}, G. and {Rain{\`o}}, S. and {Rando}, R. and {Razzano}, M. and {Razzaque}, S. and {Reimer}, A. and {Reimer}, O. and {Remy}, Q. and {Reposeur}, T. and {Romani}, R.~W. and {Saz Parkinson}, P.~M. and {Schinzel}, F.~K. and {Serini}, D. and {Sgr{\`o}}, C. and {Siskind}, E.~J. and {Smith}, D.~A. and {Spandre}, G. and {Spinelli}, P. and {Strong}, A.~W. and {Suson}, D.~J. and {Tajima}, H. and {Takahashi}, M.~N. and {Tak}, D. and {Thayer}, J.~B. and {Thompson}, D.~J. and {Tibaldo}, L. and {Torres}, D.~F. and {Torresi}, E. and {Valverde}, J. and {Van Klaveren}, B. and {van Zyl}, P. and {Wood}, K. and {Yassine}, M. and {Zaharijas}, G.},
        title = "{Fermi Large Area Telescope Fourth Source Catalog}",
      journal = {The Astrophysical Journal Supplement Series},
         year = 2020,
        month = mar,
       volume = {247},
       number = {1},
          eid = {33},
        pages = {33},
          doi = {10.3847/1538-4365/ab6bcb},
archivePrefix = {arXiv},
       eprint = {1902.10045},
 primaryClass = {astro-ph.HE},
       adsurl = {https://ui.adsabs.harvard.edu/abs/2020ApJS..247...33A}
}

@ARTICLE{Bellm2019,
       author = {{Bellm}, Eric C. and {Kulkarni}, Shrinivas R. and {Graham}, Matthew J. and {Dekany}, Richard and {Smith}, Roger M. and {Riddle}, Reed and {Masci}, Frank J. and {Helou}, George and {Prince}, Thomas A. and {Adams}, Scott M. and {Barbarino}, C. and {Barlow}, Tom and {Bauer}, James and {Beck}, Ron and {Belicki}, Justin and {Biswas}, Rahul and {Blagorodnova}, Nadejda and {Bodewits}, Dennis and {Bolin}, Bryce and {Brinnel}, Valery and {Brooke}, Tim and {Bue}, Brian and {Bulla}, Mattia and {Burruss}, Rick and {Cenko}, S. Bradley and {Chang}, Chan-Kao and {Connolly}, Andrew and {Coughlin}, Michael and {Cromer}, John and {Cunningham}, Virginia and {De}, Kishalay and {Delacroix}, Alex and {Desai}, Vandana and {Duev}, Dmitry A. and {Eadie}, Gwendolyn and {Farnham}, Tony L. and {Feeney}, Michael and {Feindt}, Ulrich and {Flynn}, David and {Franckowiak}, Anna and {Frederick}, S. and {Fremling}, C. and {Gal-Yam}, Avishay and {Gezari}, Suvi and {Giomi}, Matteo and {Goldstein}, Daniel A. and {Golkhou}, V. Zach and {Goobar}, Ariel and {Groom}, Steven and {Hacopians}, Eugean and {Hale}, David and {Henning}, John and {Ho}, Anna Y.~Q. and {Hover}, David and {Howell}, Justin and {Hung}, Tiara and {Huppenkothen}, Daniela and {Imel}, David and {Ip}, Wing-Huen and {Ivezi{\'c}}, {\v{Z}}eljko and {Jackson}, Edward and {Jones}, Lynne and {Juric}, Mario and {Kasliwal}, Mansi M. and {Kaspi}, S. and {Kaye}, Stephen and {Kelley}, Michael S.~P. and {Kowalski}, Marek and {Kramer}, Emily and {Kupfer}, Thomas and {Landry}, Walter and {Laher}, Russ R. and {Lee}, Chien-De and {Lin}, Hsing Wen and {Lin}, Zhong-Yi and {Lunnan}, Ragnhild and {Giomi}, Matteo and {Mahabal}, Ashish and {Mao}, Peter and {Miller}, Adam A. and {Monkewitz}, Serge and {Murphy}, Patrick and {Ngeow}, Chow-Choong and {Nordin}, Jakob and {Nugent}, Peter and {Ofek}, Eran and {Patterson}, Maria T. and {Penprase}, Bryan and {Porter}, Michael and {Rauch}, Ludwig and {Rebbapragada}, Umaa and {Reiley}, Dan and {Rigault}, Mickael and {Rodriguez}, Hector and {van Roestel}, Jan and {Rusholme}, Ben and {van Santen}, Jakob and {Schulze}, S. and {Shupe}, David L. and {Singer}, Leo P. and {Soumagnac}, Maayane T. and {Stein}, Robert and {Surace}, Jason and {Sollerman}, Jesper and {Szkody}, Paula and {Taddia}, F. and {Terek}, Scott and {Van Sistine}, Angela and {van Velzen}, Sjoert and {Vestrand}, W. Thomas and {Walters}, Richard and {Ward}, Charlotte and {Ye}, Quan-Zhi and {Yu}, Po-Chieh and {Yan}, Lin and {Zolkower}, Jeffry},
        title = "{The Zwicky Transient Facility: System Overview, Performance, and First Results}",
      journal = {Publications of the Astronomical Society of the Pacific},
         year = 2019,
        month = jan,
       volume = {131},
       number = {995},
        pages = {018002},
          doi = {10.1088/1538-3873/aaecbe},
archivePrefix = {arXiv},
       eprint = {1902.01932},
 primaryClass = {astro-ph.IM},
       adsurl = {https://ui.adsabs.harvard.edu/abs/2019PASP..131a8002B}
}

@ARTICLE{Cho2018,
       author = {{Cho}, Patricia B. and {Halpern}, Jules P. and {Bogdanov}, Slavko},
        title = "{Variable Heating and Flaring of Three Redback Millisecond Pulsar Companions}",
      journal = {The Astrophysical Journal},
         year = 2018,
        month = oct,
       volume = {866},
       number = {1},
          eid = {71},
        pages = {71},
          doi = {10.3847/1538-4357/aade92},
archivePrefix = {arXiv},
       eprint = {1809.00215},
 primaryClass = {astro-ph.HE},
       adsurl = {https://ui.adsabs.harvard.edu/abs/2018ApJ...866...71C}
}

@ARTICLE{Kyer2025,
       author = {{Kyer}, Rebecca and {Roy}, Subhroja and {Strader}, Jay and {Urquhart}, Ryan and {Molina}, Isabella and {Craig}, Peter and {Chomiuk}, Laura},
        title = "{Multiwavelength Evidence for Two New Candidate Transitional Millisecond Pulsars in the Subluminous Disk State: 4FGL J0639.1-8009 and 4FGL J1824.2+1231}",
      journal = {The Astrophysical Journal},
         year = 2025,
        month = apr,
       volume = {983},
       number = {2},
          eid = {112},
        pages = {112},
          doi = {10.3847/1538-4357/adc128},
archivePrefix = {arXiv},
       eprint = {2503.07709},
 primaryClass = {astro-ph.HE},
       adsurl = {https://ui.adsabs.harvard.edu/abs/2025ApJ...983..112K}
}

@ARTICLE{Linares2014,
       author = {{Linares}, M.},
        title = "{X-Ray States of Redback Millisecond Pulsars}",
      journal = {The Astrophysical Journal},
         year = 2014,
        month = nov,
       volume = {795},
       number = {1},
          eid = {72},
        pages = {72},
          doi = {10.1088/0004-637X/795/1/72},
archivePrefix = {arXiv},
       eprint = {1406.2384},
 primaryClass = {astro-ph.HE},
       adsurl = {https://ui.adsabs.harvard.edu/abs/2014ApJ...795...72L}
}

@ARTICLE{Finkbeiner2025,
       author = {{Finkbeiner}, Douglas P. and {Prince}, Thomas A. and {Whitebook}, Samuel E.},
        title = "{A Fast Periodicity Detection Algorithm Sensitive to Arbitrary Waveforms}",
      journal = {Publications of the Astronomical Society of the Pacific},
         year = 2025,
        month = may,
       volume = {137},
       number = {5},
          eid = {054504},
        pages = {054504},
          doi = {10.1088/1538-3873/add20c},
archivePrefix = {arXiv},
       eprint = {2502.00243},
 primaryClass = {astro-ph.IM},
       adsurl = {https://ui.adsabs.harvard.edu/abs/2025PASP..137e4504F}
}

@ARTICLE{Linares2018,
       author = {{Linares}, Manuel},
        title = "{The 26.3-h orbit and multiwavelength properties of the `redback' millisecond pulsar PSR J1306-40}",
      journal = {Monthly Notices of the Royal Astronomical Society},
         year = 2018,
        month = jan,
       volume = {473},
       number = {1},
        pages = {L50-L53},
          doi = {10.1093/mnrasl/slx153},
archivePrefix = {arXiv},
       eprint = {1707.00698},
 primaryClass = {astro-ph.HE},
       adsurl = {https://ui.adsabs.harvard.edu/abs/2018MNRAS.473L..50L}
}

@ARTICLE{Keane2018,
       author = {{Keane}, E.~F. and {Barr}, E.~D. and {Jameson}, A. and {Morello}, V. and {Caleb}, M. and {Bhandari}, S. and {Petroff}, E. and {Possenti}, A. and {Burgay}, M. and {Tiburzi}, C. and {Bailes}, M. and {Bhat}, N.~D.~R. and {Burke-Spolaor}, S. and {Eatough}, R.~P. and {Flynn}, C. and {Jankowski}, F. and {Johnston}, S. and {Kramer}, M. and {Levin}, L. and {Ng}, C. and {van Straten}, W. and {Krishnan}, V. Venkatraman},
        title = "{The SUrvey for Pulsars and Extragalactic Radio Bursts - I. Survey description and overview}",
      journal = {Monthly Notices of the Royal Astronomical Society},
         year = 2018,
        month = jan,
       volume = {473},
       number = {1},
        pages = {116-135},
          doi = {10.1093/mnras/stx2126},
archivePrefix = {arXiv},
       eprint = {1706.04459},
 primaryClass = {astro-ph.IM},
       adsurl = {https://ui.adsabs.harvard.edu/abs/2018MNRAS.473..116K}
}

@ARTICLE{Gomel2021,
       author = {{Gomel}, Roy and {Faigler}, Simchon and {Mazeh}, Tsevi},
        title = "{Search for dormant black holes in ellipsoidal variables I. Revisiting the expected amplitudes of the photometric modulation}",
      journal = {Monthly Notices of the Royal Astronomical Society},
         year = 2021,
        month = feb,
       volume = {501},
       number = {2},
        pages = {2822-2832},
          doi = {10.1093/mnras/staa3305},
archivePrefix = {arXiv},
       eprint = {2008.11209},
 primaryClass = {astro-ph.SR},
       adsurl = {https://ui.adsabs.harvard.edu/abs/2021MNRAS.501.2822G}
}

@ARTICLE{Bhattacharya1991,
       author = {{Bhattacharya}, D. and {van den Heuvel}, E.~P.~J.},
        title = "{Formation and evolution of binary and millisecond radio pulsars}",
      journal = {Physics Reports},
         year = 1991,
        month = jan,
       volume = {203},
       number = {1-2},
        pages = {1-124},
          doi = {10.1016/0370-1573(91)90064-S},
       adsurl = {https://ui.adsabs.harvard.edu/abs/1991PhR...203....1B}
}

@ARTICLE{Strader2025,
       author = {{Strader}, Jay and {Ray}, Paul S. and {Urquhart}, Ryan and {Swihart}, Samuel J. and {Chomiuk}, Laura and {Aydi}, Elias and {Bellm}, Eric C. and {Dage}, Kristen C. and {DeCesar}, Megan E. and {Deneva}, Julia S. and {McLaughlin}, Maura A. and {Molina}, Isabella and {Panurach}, Teresa and {Sokolovsky}, Kirill V.},
        title = "{PSR J1947{\ensuremath{-}}1120: A New Huntsman Millisecond Pulsar Binary}",
      journal = {The Astrophysical Journal},
         year = 2025,
        month = feb,
       volume = {980},
       number = {1},
          eid = {124},
        pages = {124},
          doi = {10.3847/1538-4357/ada897},
archivePrefix = {arXiv},
       eprint = {2501.05509},
 primaryClass = {astro-ph.HE},
       adsurl = {https://ui.adsabs.harvard.edu/abs/2025ApJ...980..124S}
}

@ARTICLE{Romani2026,
       author = {{Romani}, Roger W. and {Beleznay}, Maya and {Filippenko}, Alexei V. and {Brink}, Thomas G. and {Zheng}, WeiKang},
        title = "{PSR J0952-0607: Tightening a Record-high Neutron Star Mass}",
      journal = {The Astrophysical Journal},
         year = 2026,
        month = jan,
       volume = {996},
       number = {1},
          eid = {101},
        pages = {101},
          doi = {10.3847/1538-4357/ae28c5},
archivePrefix = {arXiv},
       eprint = {2512.05099},
 primaryClass = {astro-ph.HE},
       adsurl = {https://ui.adsabs.harvard.edu/abs/2026ApJ...996..101R}
}

@ARTICLE{Clark2023,
       author = {{Clark}, C.~J. and {Kerr}, M. and {Barr}, E.~D. and {Bhattacharyya}, B. and {Breton}, R.~P. and {Bruel}, P. and {Camilo}, F. and {Chen}, W. and {Cognard}, I. and {Cromartie}, H.~T. and {Deneva}, J. and {Dhillon}, V.~S. and {Guillemot}, L. and {Kennedy}, M.~R. and {Kramer}, M. and {Lyne}, A.~G. and {Mata S{\'a}nchez}, D. and {Nieder}, L. and {Phillips}, C. and {Ransom}, S.~M. and {Ray}, P.~S. and {Roberts}, M.~S.~E. and {Roy}, J. and {Smith}, D.~A. and {Spiewak}, R. and {Stappers}, B.~W. and {Tabassum}, S. and {Theureau}, G. and {Voisin}, G.},
        title = "{Neutron star mass estimates from gamma-ray eclipses in spider millisecond pulsar binaries}",
      journal = {Nature Astronomy},
         year = 2023,
        month = apr,
       volume = {7},
        pages = {451-462},
          doi = {10.1038/s41550-022-01874-x},
archivePrefix = {arXiv},
       eprint = {2301.10995},
 primaryClass = {astro-ph.HE},
       adsurl = {https://ui.adsabs.harvard.edu/abs/2023NatAs...7..451C}
}

@ARTICLE{Hessels2006,
       author = {{Hessels}, Jason W.~T. and {Ransom}, Scott M. and {Stairs}, Ingrid H. and {Freire}, Paulo C.~C. and {Kaspi}, Victoria M. and {Camilo}, Fernando},
        title = "{A Radio Pulsar Spinning at 716 Hz}",
      journal = {Science},
         year = 2006,
        month = mar,
       volume = {311},
       number = {5769},
        pages = {1901-1904},
          doi = {10.1126/science.1123430},
archivePrefix = {arXiv},
       eprint = {astro-ph/0601337},
 primaryClass = {astro-ph},
       adsurl = {https://ui.adsabs.harvard.edu/abs/2006Sci...311.1901H}
}

@ARTICLE{Koljonen2025,
       author = {{Koljonen}, Karri I.~I. and {Linares}, Manuel},
        title = "{SpiderCat: A Catalog of Compact Binary Millisecond Pulsars}",
      journal = {The Astrophysical Journal},
         year = 2025,
        month = nov,
       volume = {994},
       number = {1},
          eid = {8},
        pages = {8},
          doi = {10.3847/1538-4357/ae08a5},
archivePrefix = {arXiv},
       eprint = {2505.11691},
 primaryClass = {astro-ph.HE},
       adsurl = {https://ui.adsabs.harvard.edu/abs/2025ApJ...994....8K}
}

@ARTICLE{Chen2013,
       author = {{Chen}, Hai-Liang and {Chen}, Xuefei and {Tauris}, Thomas M. and {Han}, Zhanwen},
        title = "{Formation of Black Widows and Redbacks{\textemdash}Two Distinct Populations of Eclipsing Binary Millisecond Pulsars}",
      journal = {The Astrophysical Journal},
         year = 2013,
        month = sep,
       volume = {775},
       number = {1},
          eid = {27},
        pages = {27},
          doi = {10.1088/0004-637X/775/1/27},
archivePrefix = {arXiv},
       eprint = {1308.4107},
 primaryClass = {astro-ph.SR},
       adsurl = {https://ui.adsabs.harvard.edu/abs/2013ApJ...775...27C}
}

@ARTICLE{Benvenuto2014,
       author = {{Benvenuto}, O.~G. and {De Vito}, M.~A. and {Horvath}, J.~E.},
        title = "{Understanding the Evolution of Close Binary Systems with Radio Pulsars}",
      journal = {The Astrophysical Journal},
         year = 2014,
        month = may,
       volume = {786},
       number = {1},
          eid = {L7},
        pages = {L7},
          doi = {10.1088/2041-8205/786/1/L7},
archivePrefix = {arXiv},
       eprint = {1402.7338},
 primaryClass = {astro-ph.SR},
       adsurl = {https://ui.adsabs.harvard.edu/abs/2014ApJ...786L...7B}
}

@ARTICLE{Ginzburg2021,
       author = {{Ginzburg}, Sivan and {Quataert}, Eliot},
        title = "{Black widow formation by pulsar irradiation and sustained magnetic braking}",
      journal = {Monthly Notices of the Royal Astronomical Society},
         year = 2021,
        month = jan,
       volume = {500},
       number = {2},
        pages = {1592-1603},
          doi = {10.1093/mnras/staa3358},
archivePrefix = {arXiv},
       eprint = {2008.06506},
 primaryClass = {astro-ph.HE},
       adsurl = {https://ui.adsabs.harvard.edu/abs/2021MNRAS.500.1592G}
}

@ARTICLE{Bellm2016,
       author = {{Bellm}, Eric C. and {Kaplan}, David L. and {Breton}, Rene P. and {Phinney}, E. Sterl and {Bhalerao}, Varun B. and {Camilo}, Fernando and {Dahal}, Sumit and {Djorgovski}, S.~G. and {Drake}, Andrew J. and {Hessels}, J.~W.~T. and {Laher}, Russ R. and {Levitan}, David B. and {Lewis}, Fraser and {Mahabal}, Ashish A. and {Ofek}, Eran O. and {Prince}, Thomas A. and {Ransom}, Scott M. and {Roberts}, Mallory S.~E. and {Russell}, David M. and {Sesar}, Branimir and {Surace}, Jason A. and {Tang}, Sumin},
        title = "{Properties and Evolution of the Redback Millisecond Pulsar Binary PSR J2129-0429}",
      journal = {The Astrophysical Journal},
         year = 2016,
        month = jan,
       volume = {816},
       number = {2},
          eid = {74},
        pages = {74},
          doi = {10.3847/0004-637X/816/2/74},
archivePrefix = {arXiv},
       eprint = {1510.00721},
 primaryClass = {astro-ph.SR},
       adsurl = {https://ui.adsabs.harvard.edu/abs/2016ApJ...816...74B}
}

@ARTICLE{Misra2025,
       author = {{Misra}, Devina and {Linares}, Manuel and {Ye}, Claire S.},
        title = "{Investigating cannibalistic millisecond pulsar binaries using MESA: New constraints from pulsar spin and mass evolution}",
      journal = {Astronomy and Astrophysics},
         year = 2025,
        month = jan,
       volume = {693},
          eid = {A314},
        pages = {A314},
          doi = {10.1051/0004-6361/202452035},
archivePrefix = {arXiv},
       eprint = {2408.16048},
 primaryClass = {astro-ph.HE},
       adsurl = {https://ui.adsabs.harvard.edu/abs/2025A&A...693A.314M}
}

@ARTICLE{Sautron2026,
       author = {{Sautron}, Matt{\'e}o and {P{\'e}tri}, J{\'e}r{\^o}me and {Mitra}, Dipanjan and {Dupuy-Junet}, Ad{\'e}lie and {Pietrin}, Marie-Elo{\"\i}se},
        title = "{Born to be recycled: A comprehensive population synthesis of the Galactic millisecond pulsars}",
      journal = {Astronomy and Astrophysics},
         year = 2026,
        month = mar,
       volume = {707},
          eid = {A257},
        pages = {A257},
          doi = {10.1051/0004-6361/202557739},
archivePrefix = {arXiv},
       eprint = {2510.15661},
 primaryClass = {astro-ph.HE},
       adsurl = {https://ui.adsabs.harvard.edu/abs/2026A&A...707A.257S}
}

@ARTICLE{Breton2013,
       author = {{Breton}, R.~P. and {van Kerkwijk}, M.~H. and {Roberts}, M.~S.~E. and {Hessels}, J.~W.~T. and {Camilo}, F. and {McLaughlin}, M.~A. and {Ransom}, S.~M. and {Ray}, P.~S. and {Stairs}, I.~H.},
        title = "{Discovery of the Optical Counterparts to Four Energetic Fermi Millisecond Pulsars}",
      journal = {The Astrophysical Journal},
         year = 2013,
        month = jun,
       volume = {769},
       number = {2},
          eid = {108},
        pages = {108},
          doi = {10.1088/0004-637X/769/2/108},
archivePrefix = {arXiv},
       eprint = {1302.1790},
 primaryClass = {astro-ph.HE},
       adsurl = {https://ui.adsabs.harvard.edu/abs/2013ApJ...769..108B}
}

@ARTICLE{Roberts2014,
       author = {{Roberts}, M.~S.~E. and {Mclaughlin}, M.~A. and {Gentile}, P. and {Aliu}, E. and {Hessels}, J.~W.~T. and {Ransom}, S.~M. and {Ray}, P.~S.},
        title = "{Intrabinary shock emission from ``black widows`` and ``redbacks''}",
      journal = {Astronomische Nachrichten},
         year = 2014,
        month = mar,
       volume = {335},
       number = {3},
        pages = {313-317},
          doi = {10.1002/asna.201312038},
archivePrefix = {arXiv},
       eprint = {1402.5507},
 primaryClass = {astro-ph.HE},
       adsurl = {https://ui.adsabs.harvard.edu/abs/2014AN....335..313R}
}

@ARTICLE{1991ApJ...379L..69T,
       author = {{Tavani}, Marco},
        title = "{``Hidden'' Millisecond Pulsars}",
      journal = {\apjl},
         year = 1991,
        month = oct,
       volume = {379},
        pages = {L69},
          doi = {10.1086/186156},
       adsurl = {https://ui.adsabs.harvard.edu/abs/1991ApJ...379L..69T}
}

@INPROCEEDINGS{Papitto2022,
       author = {{Papitto}, Alessandro and {de Martino}, Domitilla},
        title = "{Transitional Millisecond Pulsars}",
    booktitle = {Astrophysics and Space Science Library},
         year = 2022,
       editor = {{Bhattacharyya}, Sudip and {Papitto}, Alessandro and {Bhattacharya}, Dipankar},
       series = {Astrophysics and Space Science Library},
       volume = {465},
        month = jan,
        pages = {157-200},
          doi = {10.1007/978-3-030-85198-9_6},
archivePrefix = {arXiv},
       eprint = {2010.09060},
 primaryClass = {astro-ph.HE},
       adsurl = {https://ui.adsabs.harvard.edu/abs/2022ASSL..465..157P}
}

@ARTICLE{Vilhu1987,
       author = {{Vilhu}, Osmi and {Walter}, Frederick M.},
        title = "{Chromospheric-Coronal Activity at Saturated Levels}",
      journal = {The Astrophysical Journal},
         year = 1987,
        month = oct,
       volume = {321},
        pages = {958},
          doi = {10.1086/165689},
       adsurl = {https://ui.adsabs.harvard.edu/abs/1987ApJ...321..958V}
}

\onecolumn
\appendix
\section{Tables of Photometry and Radial Velocities}\label{Appendix}
\setcounter{table}{0}
\renewcommand{\thetable}{A\arabic{table}}

\begin{longtable}{crc}
\caption{SOAR Photometry for 4FGL J1020.4--5314}
\label{table:j1020_phot}\\
\hline
BMJD$^{a}$ & $i$ & unc. \\
(d) & (mag) & (mag) \\
\hline
\endfirsthead
\multicolumn{3}{c}{\textit{Table~\ref{table:j1020_phot} -- continued from previous page}}\\
\hline
BMJD$^{a}$ & $i$ & unc. \\
(d) & (mag) & (mag) \\
\hline
\endhead
\hline
\multicolumn{3}{r}{\textit{continued on next page}}\\
\endfoot
\hline
\multicolumn{3}{l}{$^{a}$\,Barycentric Modified Julian Date.}\\
\endlastfoot
59739.9785613 & 19.144 & 0.011  \\
59739.9809149 & 19.146 & 0.011  \\
59739.9830832 & 19.169 & 0.012  \\
59739.9852514 & 19.180 & 0.011  \\
59739.9874201 & 19.160 & 0.011  \\
59739.9895884 & 19.168 & 0.011  \\
59739.9918149 & 19.181 & 0.012  \\
59739.9939839 & 19.167 & 0.012  \\
59739.9961523 & 19.207 & 0.011  \\
59739.9983211 & 19.203 & 0.012  \\
59740.0004895 & 19.210 & 0.012  \\
59740.0026584 & 19.207 & 0.013  \\
59740.0049519 & 19.202 & 0.012  \\
59740.0071209 & 19.204 & 0.013  \\
59740.0092899 & 19.228 & 0.014  \\
59740.0114583 & 19.240 & 0.017  \\
59740.0136266 & 19.235 & 0.015  \\
59740.0157952 & 19.237 & 0.014  \\
59740.0179636 & 19.242 & 0.012  \\
59740.0201320 & 19.263 & 0.013  \\
59740.0223009 & 19.283 & 0.014  \\
59740.0244692 & 19.309 & 0.013  \\
59740.0266858 & 19.332 & 0.016  \\
59740.0288546 & 19.344 & 0.014  \\
59740.0310227 & 19.416 & 0.014  \\
59740.0331910 & 19.456 & 0.016  \\
59740.0353594 & 19.483 & 0.017  \\
59740.0375286 & 19.516 & 0.017  \\
59740.0396972 & 19.557 & 0.017  \\
59740.0418668 & 19.592 & 0.016  \\
59740.0440359 & 19.639 & 0.019  \\
59740.0462039 & 19.705 & 0.025  \\
59740.0484421 & 19.714 & 0.029  \\
59740.0506107 & 19.731 & 0.029  \\
59740.0527794 & 19.746 & 0.029  \\
59740.0549483 & 19.817 & 0.051  \\
59740.0571162 & 19.825 & 0.042  \\
59740.0592844 & 19.804 & 0.027  \\
59740.0614538 & 19.809 & 0.027  \\
59740.0636223 & 19.821 & 0.030  \\
59740.0657910 & 19.845 & 0.031  \\
59740.0679598 & 19.773 & 0.027  \\
59740.0702354 & 19.751 & 0.030  \\
59740.0724039 & 19.802 & 0.042  \\
59740.0745722 & 19.695 & 0.072  \\
59740.0767404 & 19.756 & 0.050  \\
59740.0789086 & 19.690 & 0.038  \\
59740.0810771 & 19.660 & 0.025  \\
59740.0832459 & 19.593 & 0.019  \\
59740.0854146 & 19.561 & 0.018  \\
59740.0875831 & 19.508 & 0.016  \\
59740.0897514 & 19.465 & 0.016  \\
59740.0919418 & 19.417 & 0.016  \\
59740.0941101 & 19.413 & 0.024  \\
59740.0962785 & 19.381 & 0.033  \\
60039.0635765 & 19.357 & 0.009  \\
60039.0657449 & 19.337 & 0.008  \\
60039.0679134 & 19.342 & 0.008  \\
60039.0700821 & 19.287 & 0.009  \\
60039.0722503 & 19.269 & 0.008  \\
60039.0744659 & 19.243 & 0.008  \\
60039.0766347 & 19.251 & 0.008  \\
60039.0788034 & 19.255 & 0.007  \\
60039.0809723 & 19.253 & 0.008  \\
60039.0831407 & 19.273 & 0.008  \\
60039.0854662 & 19.250 & 0.008  \\
60039.0876349 & 19.248 & 0.008  \\
60039.0898029 & 19.261 & 0.009  \\
60039.0919715 & 19.279 & 0.008  \\
60039.0941398 & 19.249 & 0.008  \\
60039.0966569 & 19.293 & 0.009  \\
60039.0988251 & 19.311 & 0.009  \\
60039.1009940 & 19.310 & 0.009  \\
60039.1031623 & 19.311 & 0.008  \\
60039.1053311 & 19.305 & 0.008  \\
60039.1080300 & 19.328 & 0.009  \\
60039.1101988 & 19.336 & 0.009  \\
60039.1123675 & 19.352 & 0.009  \\
60039.1145361 & 19.352 & 0.009  \\
60039.1167046 & 19.315 & 0.009  \\
60039.1196237 & 19.322 & 0.010  \\
60039.1217926 & 19.302 & 0.009  \\
60039.1239613 & 19.317 & 0.010  \\
60039.1261302 & 19.303 & 0.009  \\
60039.1282989 & 19.301 & 0.010  \\
60039.1314087 & 19.300 & 0.009  \\
60039.1335777 & 19.281 & 0.010  \\
60039.1357459 & 19.317 & 0.009  \\
60039.1379145 & 19.308 & 0.010  \\
60039.1400832 & 19.299 & 0.009  \\
60039.1429737 & 19.336 & 0.010  \\
60039.1451421 & 19.332 & 0.009  \\
60039.1473109 & 19.350 & 0.010  \\
60039.1494796 & 19.376 & 0.009  \\
60039.1516482 & 19.396 & 0.011  \\
60039.1538677 & 19.415 & 0.011  \\
60039.1560359 & 19.452 & 0.012  \\
60039.1582052 & 19.447 & 0.012  \\
60039.1603739 & 19.493 & 0.012  \\
60039.1625426 & 19.500 & 0.012  \\
60039.1655419 & 19.565 & 0.012  \\
60039.1677108 & 19.577 & 0.012  \\
60039.1698794 & 19.585 & 0.012  \\
60039.1720479 & 19.613 & 0.013  \\
60039.1742166 & 19.651 & 0.013  \\
60039.1764633 & 19.640 & 0.013  \\
60039.1786319 & 19.667 & 0.014  \\
60039.1808004 & 19.667 & 0.012  \\
60039.1829685 & 19.685 & 0.011  \\
60039.1851370 & 19.661 & 0.011  \\
60039.1873194 & 19.674 & 0.011  \\
60039.1894882 & 19.674 & 0.011  \\
60039.1916572 & 19.639 & 0.011  \\
60039.1938257 & 19.603 & 0.010  \\
60039.1959950 & 19.603 & 0.010  \\
60039.1982706 & 19.565 & 0.011  \\
60039.2004395 & 19.550 & 0.010  \\
60039.2026078 & 19.525 & 0.010  \\
60039.2047767 & 19.506 & 0.010  \\
60039.2069454 & 19.464 & 0.010  \\
60039.2091374 & 19.448 & 0.012  \\
60039.2113062 & 19.418 & 0.012  \\
60039.2134749 & 19.373 & 0.011  \\
60039.2156439 & 19.346 & 0.010  \\
60039.2178125 & 19.322 & 0.011  \\
60039.2200090 & 19.295 & 0.010  \\
60039.2221773 & 19.281 & 0.010  \\
60039.2243460 & 19.274 & 0.010  \\
60039.2265150 & 19.227 & 0.009  \\
60039.2286834 & 19.238 & 0.009  \\
60039.2308653 & 19.224 & 0.008  \\
60039.2330343 & 19.223 & 0.009  \\
60039.2352030 & 19.225 & 0.008  \\
60039.2373719 & 19.218 & 0.008  \\
60039.2395406 & 19.220 & 0.008  \\
60039.2417692 & 19.232 & 0.008  \\
60039.2439379 & 19.248 & 0.009  \\
60039.2461067 & 19.230 & 0.008  \\
60039.2482754 & 19.240 & 0.009  \\
60039.2504442 & 19.247 & 0.008  \\
60039.2529801 & 19.246 & 0.009  \\
60039.2551486 & 19.257 & 0.009  \\
60039.2573174 & 19.259 & 0.009  \\
60039.2594861 & 19.265 & 0.009  \\
60039.2616545 & 19.289 & 0.009  \\
60039.2638693 & 19.284 & 0.009  \\
60039.2660377 & 19.283 & 0.009  \\
60039.2682062 & 19.261 & 0.011  \\
60039.2703751 & 19.270 & 0.010  \\
60039.2725440 & 19.295 & 0.009  \\
60039.2757365 & 19.270 & 0.009  \\
60039.2779052 & 19.261 & 0.009  \\
60039.2800742 & 19.268 & 0.008  \\
60039.2822428 & 19.255 & 0.009  \\
60039.2844112 & 19.289 & 0.009  \\
60039.2865898 & 19.269 & 0.009  \\
60039.2887588 & 19.274 & 0.010  \\
60039.2909273 & 19.280 & 0.010  \\
60039.2930958 & 19.316 & 0.010  \\
60039.2952650 & 19.321 & 0.010  \\
60341.2976545 & 19.608 & 0.055  \\
60341.2993981 & 19.546 & 0.018  \\
60341.3015668 & 19.562 & 0.017  \\
60341.3037355 & 19.588 & 0.018  \\
60341.3059044 & 19.587 & 0.018  \\
60341.3080740 & 19.607 & 0.017  \\
60341.3102432 & 19.609 & 0.018  \\
60341.3124114 & 19.689 & 0.019  \\
60341.3145806 & 19.682 & 0.020  \\
60341.3167497 & 19.687 & 0.018  \\
60341.3189185 & 19.715 & 0.022  \\
60341.3212245 & 19.704 & 0.019  \\
60341.3233933 & 19.706 & 0.018  \\
60341.3255622 & 19.674 & 0.018  \\
60341.3277309 & 19.689 & 0.019  \\
60341.3299002 & 19.687 & 0.019  \\
60341.3320692 & 19.686 & 0.019  \\
60341.3342379 & 19.676 & 0.018  \\
60341.3364063 & 19.630 & 0.017  \\
60341.3385753 & 19.602 & 0.017  \\
60341.3407440 & 19.594 & 0.017  \\
60341.3429608 & 19.580 & 0.016  \\
60341.3451307 & 19.537 & 0.018  \\
60341.3472997 & 19.493 & 0.016  \\
60341.3494687 & 19.502 & 0.016  \\
60341.3516376 & 19.441 & 0.016  \\
60341.3538061 & 19.420 & 0.014  \\
60341.3559746 & 19.384 & 0.013  \\
60341.3581440 & 19.362 & 0.015  \\
\end{longtable}
\twocolumn

\begin{table}
\centering
\caption{Radial Velocities for 4FGL J1020.4--5314}
\label{table:rv}
\begin{tabular}{crcrc}
\hline
BMJD$^{a}$ & radial velocity$^{b}$ & unc. & H$\alpha$ vel. & unc.  \\
(d) & (km\,s$^{-1}$) & (km\,s$^{-1}$) & (km\,s$^{-1}$) & (km\,s$^{-1}$) \\
\hline
59679.0079412 & 407.1 & 15.2 & --9.5 & 24.2 \\
59679.0262893 & 286.5 & 16.5 & --118.8 & 30.4 \\
59680.0544190 & 418.9 & 19.9 & 20.3 & 33.5 \\
59680.0721532 & 386.4 & 17.9 & --34.4 & 42.8 \\
59680.0908298 & 232.7 & 20.5 & --150.4 & 77.0 \\
59680.1484756 & --289.1 & 34.8 & 188.6 & 30.0 \\
59680.1678371 & --40.7 & 50.3 & 195.3 & 20.8 \\
59680.2040237 & 475.6 & 25.9 & --35.6 & 35.7 \\
59691.0157913 & --249.5 & 31.3 & 158.2 & 45.9 \\
59722.1405976 & 465.7 & 23.0 & --7.6 & 40.0 \\
59724.0055201 & --56.6 & 16.8 & 4.8 & 44.4 \\
59724.0234332 & --202.1 & 23.8 & 250.7 & 38.3 \\
60121.9888259 & 42.6 & 30.6 & 198.7 & 48.1 \\
60268.2589247 & --189.9 & 16.5 & 125.2 & 23.7 \\
60268.2764023 & --321.4 & 26.8 & 4.9 & 21.0 \\
60302.2527134 & 87.1 & 52.1 & 103.2 & 23.4 \\
60302.2705340 & 399.4 & 42.0 & 43.6 & 38.1 \\
60302.2938757 & 408.3 & 37.2 & --35.2 & 38.7 \\
60319.2053863 & 306.6 & 17.5 & --30.6 & 23.8 \\
60320.2383741 & 480.4 & 24.5 & --65.7 & 27.4 \\
60320.2561700 & 369.2 & 14.5 & 8.8 & 33.8 \\
60320.2967186 & --162.3 & 15.4 & 93.0 & 21.9 \\
60320.3203325 & --345.7 & 18.0 & 107.6 & 22.3 \\
60320.3378109 & --86.2 & 24.7 & 100.8 & 16.0 \\
60320.3610728 & 230.3 & 59.3 & 57.6 & 58.9 \\
60331.2319647 & 401.8 & 62.8 & --104.0 & 26.9 \\
60331.2591216 & 437.4 & 44.2 & --76.7 & 36.3 \\
60331.2765990 & 256.1 & 27.1 & --62.5 & 38.0 \\
\hline
\end{tabular}
\\[2pt]
{\raggedright $^{a}$\,Barycentric Modified Julian Date.\par}
{\raggedright $^{b}$\,Absorption-line radial velocity.\par}
\end{table}

\begin{table}
\centering
\caption{Absorption-Line Radial Velocities for 4FGL J1408.7--3747}
\label{table:j1408rv}
\begin{tabular}{crc}
\hline
BMJD$^{a}$ & radial velocity & unc. \\
(d) & (km\,s$^{-1}$) & (km\,s$^{-1}$) \\
\hline
59679.1640220 & --51.6 & 11.7 \\
59680.2942122 & 34.8 & 10.7 \\
59680.3083819 & --27.3 & 9.1 \\
60494.0502318 & --58.6 & 11.7 \\
60494.0677072 & --95.6 & 9.0 \\
60515.0259429 & 17.8 & 21.3 \\
60749.3352741 & 25.6 & 24.4 \\
60749.3458075 & --0.9 & 23.4 \\
60760.2433634 & 267.1 & 15.8 \\
60760.2538965 & 246.1 & 15.1 \\
60792.0960024 & 260.3 & 16.1 \\
60792.1065349 & 280.0 & 17.3 \\
60792.1196034 & 355.9 & 16.8 \\
60792.1301354 & 358.4 & 15.0 \\
60812.0476667 & --56.4 & 15.0 \\
60812.0585106 & --98.4 & 14.3 \\
60812.0750539 & --114.9 & 13.1 \\
60812.0855855 & --115.6 & 13.1 \\
60812.1059106 & --94.3 & 15.9 \\
60812.1164424 & --47.0 & 15.7 \\
60812.1575481 & 188.0 & 15.8 \\
60812.1680797 & 243.1 & 18.3 \\
60812.1809921 & 293.1 & 19.5 \\
60812.1915240 & 327.0 & 17.7 \\
60812.2072336 & 387.5 & 15.9 \\
60812.2177658 & 408.3 & 15.1 \\
60812.2307223 & 413.1 & 14.5 \\
60812.2412539 & 394.9 & 14.2 \\
60823.0185614 & --118.5 & 17.9 \\
60823.0290940 & --133.5 & 17.4 \\
60823.0428028 & --114.8 & 16.0 \\
60823.0533349 & --112.7 & 14.9 \\
60823.0716334 & --62.6 & 18.5 \\
60823.0821654 & --25.9 & 17.4 \\
60823.1235849 & 208.8 & 18.2 \\
60823.1341172 & 281.6 & 19.8 \\
60854.0422018 & --66.8 & 18.4 \\
60854.0528631 & --110.8 & 20.5 \\
60854.0674544 & --128.8 & 24.2 \\
60854.0780978 & --115.9 & 34.4 \\
61076.3737394 & 412.0 & 7.9 \\
\hline
\end{tabular}
\\[2pt]
{\raggedright $^{a}$\,Barycentric Modified Julian Date.\par}
\end{table}

\onecolumn
\begin{longtable}{crcrc}
\caption{Radial Velocities for 4FGL J1429.8--0739}
\label{table:j1429rv}\\
\hline
BMJD$^{a}$ & radial velocity$^{b}$ & unc. & H$\alpha$ vel. & unc. \\
(d) & (km\,s$^{-1}$) & (km\,s$^{-1}$) \\
\hline
\endfirsthead
\multicolumn{3}{c}{\textit{Table~\ref{table:j1429rv} -- continued from previous page}}\\
\hline
BMJD$^{a}$ & radial velocity & unc. \\
(d) & (km\,s$^{-1}$) & (km\,s$^{-1}$) \\
\hline
\endhead
\hline
\multicolumn{3}{r}{\textit{continued on next page}}\\
\endfoot
\hline
\multicolumn{3}{l}{$^{a}$\,Barycentric Modified Julian Date.}\\
\multicolumn{3}{l}{$^{b}$\,Absorption-line radial velocity.}\\
\endlastfoot
60737.3138638 & 152.0 & 17.2 & 223.5 & 12.2 \\
60737.3278700 & 146.3 & 18.3 & 238.4 & 7.0 \\
60760.2734606 & 169.1 & 18.1 & 68.3 & 13.3 \\
60760.2874661 & 152.1 & 20.2 & 52.4 & 9.1 \\
60760.3068190 & 114.6 & 20.0 & 22.1 & 11.2 \\
60760.3208244 & 160.4 & 25.8 & 14.6 & 17.5 \\
60792.1507578 & 159.1 & 28.5 & 81.7 & 29.3 \\
60792.1647627 & 172.6 & 24.1 & 134.1 & 15.5 \\
60792.1834527 & 146.3 & 22.4 & 127.8 & 23.0 \\
60792.1974569 & 125.5 & 22.4 & 47.4 & 28.3 \\
60792.2170939 & 112.2 & 17.5 & 135.3 & 16.9 \\
60792.2310983 & 105.7 & 18.1 & 122.0 & 11.4 \\
60792.2474656 & 87.4 & 20.4 & 34.7 & 10.4 \\
60792.2614697 & 26.6 & 20.5 & 73.6 & 25.9 \\
60812.2643613 & 37.0 & 22.3 & 76.6 & 31.7 \\
60812.2786707 & 81.6 & 21.7 & 29.9 & 15.8 \\
60823.1550398 & --147.2 & 21.6 & --123.8 & 11.4 \\
60823.1690448 & --162.1 & 21.5 & --126.8 & 12.9 \\
60823.1971894 & --158.9 & 26.0 & --118.4 & 22.0 \\
60823.2111937 & --216.2 & 23.3 & --134.6 & 10.1 \\
60823.2275990 & --199.8 & 22.6 & --129.9 & 20.9 \\
60823.2416037 & --149.7 & 28.6 & --149.6 & 14.8 \\
60823.2606718 & --250.8 & 27.8 & --121.7 & 23.8 \\
60823.2746763 & --178.7 & 27.7 & --71.0 & 23.8 \\
60881.0506689 & 119.8 & 20.3 & 58.4 & 25.3 \\
60881.0646718 & 155.2 & 20.0 & 97.0 & 17.5 \\
60881.0821120 & 146.9 & 20.9 & 110.7 & 39.0 \\
60881.0961146 & 149.0 & 21.4 & 113.5 & 12.2 \\
60881.1171566 & 190.6 & 25.3 & 21.5 & 39.0 \\
60920.9924063 & --97.0 & 30.1 & --59.3 & 18.6 \\
60921.0064095 & --196.1 & 38.3 & --115.7 & 30.6 \\
61077.3616653 & --100.9 & 17.3 & --68.5 & 8.8 \\
61077.3791435 & --100.8 & 16.7 & --79.3 & 6.7 \\
61086.2785552 & --101.0 & 19.1 & --131.3 & 11.2 \\
61086.3023494 & --150.8 & 15.7 & --129.6 & 5.4 \\
61086.3198277 & --174.2 & 15.1 & --159.0 & 8.2 \\
61086.3604726 & --165.2 & 15.2 & --130.2 & 9.9 \\
61087.2467474 & 131.5 & 19.4 & 56.1 & 19.7 \\
61087.2642254 & 155.4 & 18.2 & 73.4 & 14.5 \\
61087.2883143 & 68.0 & 15.1 & 56.6 & 12.6 \\
61087.3057928 & 78.5 & 14.6 & 59.0 & 12.9 \\
61087.3288039 & 25.6 & 14.7 & 44.8 & 10.4 \\
61087.3462821 & 40.8 & 14.7 & 32.1 & 11.7 \\
61087.3699296 & 10.5 & 14.7 & --13.1 & 11.2 \\
61087.3874078 & --3.9 & 14.9 & --15.2 & 7.1 \\
61150.0896778 & --215.0 & 19.5 & --99.3 & 14.2 \\
61150.1071552 & --173.5 & 18.3 & --124.4 & 11.7 \\
61150.1303807 & --146.9 & 15.6 & --108.9 & 11.8 \\
61150.1478576 & --133.3 & 16.1 & --100.5 & 5.8 \\
61150.1717628 & --160.4 & 22.7 & --114.6 & 11.9 \\
61150.1892397 & --156.2 & 15.5 & --92.8 & 6.4 \\
61150.2153520 & --147.4 & 15.5 & --125.1 & 8.2 \\
61150.2328296 & --137.2 & 15.3 & --105.0 & 9.3 \\
61150.2581143 & --138.8 & 15.7 & --116.0 & 10.1 \\
61150.2755914 & --94.8 & 17.1 & --65.9 & 9.0 \\
61154.1498139 & --23.1 & 19.5 & 1.4 & 8.3 \\
61154.1672910 & 5.8 & 19.2 & 9.8 & 16.4 \\
61154.1888880 & 6.5 & 16.8 & --6.7 & 15.1 \\
61154.2063647 & 5.9 & 17.7 & 35.0 & 20.6 \\
61154.2299120 & 32.6 & 16.9 & 30.5 & 22.4 \\
61154.2473893 & 20.1 & 15.9 & 18.1 & 22.1 \\
61154.2724645 & 24.3 & 16.4 & 38.0 & 17.0 \\
61154.2904202 & 63.1 & 15.9 & 31.4 & 21.8 \\

\end{longtable}
\twocolumn

%%%%%%%%%%%%%%%%%%%%%%%%%%%%%%%%%%%%%%%%%%%%%%%%%%

% Don't change these lines
\bsp	% typesetting comment
\label{lastpage}
\end{document}